\documentclass[aip,jcp,reprint,superscriptaddress,nofootinbib]{revtex4-2}

\usepackage{graphicx}

\usepackage{amsmath,amssymb,bm,mathtools}
\usepackage{dsfont} % proper double-stroke identity \mathds{1}
\usepackage{booktabs}
\usepackage{array}
\usepackage{xcolor}
\usepackage{tikz}
\usetikzlibrary{arrows.meta,positioning,shapes.geometric,calc,decorations.pathreplacing,fit,backgrounds}
\usepackage{enumitem}
\usepackage[noend]{algpseudocode}
\newcounter{algorithm}
\renewcommand{\thealgorithm}{\arabic{algorithm}}

\usepackage[version=4]{mhchem}
\usepackage{xspace}
\usepackage{hyperref}
\hypersetup{colorlinks=true, allcolors=blue!55!black}

\graphicspath{{figs/}{figures/}}
\providecommand{\citenum}{}\renewcommand{\citenum}{\onlinecite}

\newcommand{\Id}{\mathds{1}} % identity operator
\newcommand{\Rx}{\partial_{\xi}}
\newcommand{\Jx}{J^{\xi}}
\newcommand{\Kx}{K^{\xi}}
\newcommand{\hhTDA}{\textit{hh}\text{-TDA}\xspace}
\newcommand{\ppTDA}{\textit{pp}\text{-TDA}\xspace}
\newcommand{\WL}{W_{\mathrm{L}}}
\newcommand{\WR}{W_{\mathrm{R}}}
\newcommand{\gact}{g^{\mathrm{act\text{-}excl}}}
\newcommand{\Bq}{B^{Q}}
\newcommand{\Nel}{N_{\mathrm{el}}}
\newcommand{\Sz}{\hat{S}_z}
\newcommand{\aC}{a_{\mathrm{C}}}

\begin{document}

\title{Analytic Gradients and Nonadiabatic Couplings for Device-Resident
DMRG-QD-NEVPT2 Through Conical Intersections on a Consumer GPU}

\author{Rub\'en D. Guerrero}
\email{rudaguerman@gmail.com}
\affiliation{NeuroTechNet S.A.S., 1108831, Bogot\'a, Colombia}
\affiliation{Quantum and Computational Chemistry Group (QCCG),
Universidad Nacional de Colombia, Bogot\'a, Colombia}

\date{\today}

\begin{abstract}
Nonadiabatic dynamics through a conical intersection requires both static and dynamic
correlation and, at every geometry, an excited-state gradient and interstate
nonadiabatic coupling (NACME). Analytic multireference derivatives at this
level have been confined to datacenter-scale hardware. We report device-resident
\mbox{DMRG-QD-NEVPT2} with analytic gradients and NACMEs within the $8$\,GB of a
consumer GPU: a density-matrix-renormalization-group (DMRG) active-space reference
supplies the static correlation, and quasi-degenerate SC-NEVPT2 dresses it
through a multi-state effective Hamiltonian that remains valid \emph{through} a conical
intersection, where single-state perturbation theory develops a vanishing
denominator and fails. Every gradient and NACME is one reverse-mode transpose of a single
contraction graph: the DMRG sweep enters as a gauge-free differentiable node,
and all two-electron work is AO-direct through Cholesky factors. The DMRG
reference is exact within the active space ($\sim\!10^{-15}$\,Ha). The corrected adiabats close
smoothly to a real $S_0/S_1$ conical intersection with a non-vanishing, $1/\Delta
E$-divergent NACME---where linear-response TDDFT returns zero identically---and
the dressing restores the differential dynamic correlation that ionic
$\pi\pi^*$ states require. DMRG-NEVPT2 and quasi-degenerate NEVPT2 are each established;
what is new is the combined stack---multi-state and through a conical intersection, with
analytic gradients and couplings---realized device-resident on commodity hardware.\end{abstract}

\maketitle

%% =====================================================================
\section{Introduction}
\label{sec:intro}
%% =====================================================================
Photochemistry is decided at conical intersections, and following it by nonadiabatic
dynamics demands, at every nuclear geometry along every trajectory, two derivatives of the
electronic wavefunction: an excited-state \emph{gradient} to propagate the nuclei and a
\emph{nonadiabatic coupling matrix element} (NACME) to fix the rate of population transfer
between states~\cite{Yarkony1996}. In a companion paper we supplied these for the
density-functional excited-state family---the hole-hole and particle-particle Tamm--Dancoff
manifolds (\hhTDA/\ppTDA) and a CIS(D)-type double-hybrid coupling---obtained automatically
as the reverse-mode transpose of a single contraction graph and engineered to run
device-resident within the $8$\,GB of a consumer GPU~\cite{GuerreroDFTdag}. The \hhTDA/\ppTDA manifolds already reach past adiabatic linear response: built by double
electron attachment or detachment from an $(N\!\pm\!2)$ reference, they treat ground and
excited states on equal footing and thereby capture static correlation---near- and exact
degeneracies, and hence the conical intersections that adiabatic
TDDFT~\cite{Levine2006,Gozem2014} cannot---alongside the dynamic correlation supplied by the
density functional~\cite{Bannwarth2020}. Their reach, however, is that of a \emph{compact,
method-fixed} active space: the states accessible by that single double annihilation, in
practice a few frontier orbitals, not an active space one enlarges at will. Strong
correlation of larger extent---stretched-bond and polyradical manifolds, and the genuinely
\emph{doubly-excited} valence states of extended chromophores whose multiconfigurational
character spreads over many orbitals---calls instead for a static-correlation treatment that
is \emph{systematically improvable in an arbitrarily large active space}, which is what the
density matrix renormalization group supplies.

The density matrix renormalization group (DMRG)~\cite{White1992,White1993,Chan2011,Baiardi2020}
is the established route to those states: a variational, systematically improvable
approximation to the full configuration-interaction (FCI) wavefunction in a chosen active
space, controlled by a single bond dimension and exact in the large-bond-dimension limit.
Analytic nuclear gradients for DMRG-based methods are a decade old and under active
development, contrary to a common impression: Liu, Kurashige, Yanai and
Morokuma~\cite{LiuKurashige2013} and Nakatani, Yanai and Kurashige~\cite{Nakatani2017} gave
analytic DMRG-CASSCF (and CASPT2) gradients through a reduced-density-matrix response
interface; Hu and Chan~\cite{HuChan2015} optimized DMRG excited-state geometries; Iino,
Shiozaki and Yanai~\cite{Iino2023} derived the coupled-perturbed DMRG equations for
state-averaged DMRG-CASSCF analytic gradients; and, decisively for the present work,
Freitag, Ma, Baiardi, Knecht and Reiher~\cite{Freitag2019} reported analytic gradients
\emph{and} nonadiabatic couplings for state-averaged DMRG-SCF and located a conical
intersection with them. Each of these contributions delivers a gradient (and, in
Ref.~\cite{Freitag2019}, a coupling) at the \emph{self-consistent-field} level: the DMRG
active space plus orbital optimization, with no dynamic-correlation correction entering the
derivative. Where dynamic correlation does enter a multireference derivative---the internally
contracted NEVPT2 analytic gradients and quasidegenerate interstate couplings of
Park~\cite{Park2019}, or the multiconfiguration pair-density-functional (MC-PDFT/DMRG-PDFT)
gradients of the Gagliardi--Truhlar school~\cite{Sand2017,DMRGPDFT2019}---it has been realized
conventionally, at datacenter scale, and (for the perturbative interstate couplings) on modest
CASSCF active spaces. Our contribution is not the \emph{existence} of a quasidegenerate-NEVPT2
gradient but its realization: strongly-contracted QD-NEVPT2 built on a DMRG active-space
reference, its analytic gradients and interstate NACMEs obtained as one reverse-mode transpose
of a device-resident energy graph, and demonstrated smooth through a real conical intersection
inside the $8$\,GB of a consumer GPU.

We report precisely that realization: DMRG-QD-NEVPT2 with analytic gradients and interstate
NACMEs, device-resident on a consumer GPU. We are explicit about the delimitation: we do
\emph{not} claim the first DMRG gradient (state-averaged DMRG-SCF gradients and NACMEs are
established~\cite{Freitag2019,Iino2023}) nor the first NEVPT2 analytic derivative
(Park~\cite{Park2019}); the novelty, after an explicit literature review, is the
\emph{device-resident} DMRG-QD-NEVPT2 first derivative---gradients and interstate
couplings---formulated as a contraction-DAG transpose and run end to end on consumer hardware
through a conical intersection.

Three features distinguish this realization and constitute the substance of the paper. \emph{First}, the dynamic-correlation term is a \emph{quasi-degenerate},
multi-state strongly-contracted NEVPT2: a Hermitized quasi-degenerate NEVPT2 effective Hamiltonian over the fixed
model space (Sec.~\ref{sec:dmp2}) whose diagonalization keeps the adiabatic gap non-negative
[Eq.~\eqref{eq:gap}] and the description valid through a conical intersection, precisely where a
single-state second-order treatment develops a spurious vanishing denominator and a negative
gap; in the single-determinant limit, its diagonal reduces to active-excluded MP2---the bridge
that eliminates double-counting with respect to the hybrid Kohn--Sham reference. \emph{Second},
the entire derivative---including the DMRG sweep---is obtained tape-free as one reverse-mode
transpose of the forward energy graph (Sec.~\ref{sec:transpose}): the DMRG optimization is
carried as a single differentiable node whose adjoint is an implicit fixed-point solve, and
the tangent is taken in the gauge-free Haegeman form~\cite{Haegeman2016,Haegeman2011} so
that the gradient is exact at \emph{every} bond dimension rather than only at full bond---a
distinction with teeth, because a truncated matrix-product state is energy-converged yet not
tangent-stationary (Sec.~\ref{sec:nonstat}). \emph{Third}, the whole calculation---the DMRG
sweep, the reduced-density matrices, the quasi-degenerate NEVPT2 correction, the response, and
the \hhTDA/\ppTDA manifolds served by the same engine---runs device-resident
within the $8$\,GB of a consumer GPU, whereas efficient DMRG and its gradients have
been datacenter-scale~\cite{Zhai2023}. The DFT hh/pp-TDA and CIS(D) results of paper
1~\cite{GuerreroDFTdag} are cited as prior art and are not re-derived; \hhTDA/\ppTDA appear
here as members of the unified excited-state family that the multireference engine serves,
against which DMRG-QD-NEVPT2 is compared.

Each ingredient rests on a mature literature, which we engage directly so that the
contribution is not overstated. DMRG and its quantum-chemical formulation are those of White, Chan, Legeza,
Reiher, Yanai and co-workers~\cite{White1992,Chan2011,Legeza2003,Keller2015,Baiardi2020};
the entanglement diagnostics (single-orbital entropy, mutual information, the
dynamically-adapted block state selection) are Legeza, Rissler--Noack--White, and the
Reiher-group AutoCAS line of Stein and Reiher~\cite{Legeza2003,Rissler2006,Stein2016,Stein2019};
active-space selection by atomic-valence projection is AVAS of Sayfutyarova and
Knizia~\cite{Sayfutyarova2017}. The reverse-mode view of relaxation is likewise
established---differentiable programming through electronic structure from the
Hartree--Fock proof of concept of Tamayo-Mendoza \emph{et al.}~\cite{Tamayo2018} and the
differentiable electronic-structure framework of Zhang and Chan~\cite{Zhang2022}, and automatic differentiation of
tensor-network contractions of Liao \emph{et al.}~\cite{Liao2019}. These approaches differentiate the energy with respect to
tensor or Hamiltonian parameters, for variational optimization; here we take \emph{nuclear} derivatives---gradients
and interstate NACMEs---of a dynamic-correlation-corrected method, a different object. The
open residual their union leaves, and this paper's central result, is analytic gradients and
NACMEs for DMRG-QD-NEVPT2, from one reverse-mode contraction graph, on a consumer
card.

A methodological thread runs throughout. Because the excitation energies are exact within
the active space (DMRG$=$FCI), the residual error is diagnosed, not assumed: we report a
molecule-blocked, character-balanced (class-macro) mean absolute deviation with a
bootstrapped confidence interval (Sec.~\ref{sec:results-quest}), turning the comparison
into a quantitative, reproducible statement about which state characters the present level
of theory captures---and the answer, reported untrimmed, is the physically expected one.

%% =====================================================================
\section{Theory}
\label{sec:theory}
%% =====================================================================
We develop the three pillars just introduced in turn---the energy expression
(Sec.~\ref{sec:energy}), the quasi-degenerate NEVPT2 correction (Sec.~\ref{sec:dmp2}), and the
contraction-DAG transpose that differentiates the whole construction, including the DMRG sweep,
in one reverse pass (Sec.~\ref{sec:transpose})---before turning to the DMRG-specific caveats that make the construction work in practice (Secs.~\ref{sec:asymJK}--\ref{sec:compress}).

\subsection{The DMRG-QD-NEVPT2 energy}
\label{sec:energy}
The reference is a closed-shell Kohn--Sham determinant from a range-separated or global
hybrid, from which a Cholesky decomposition (CD) of the electron-repulsion
integrals~\cite{Beebe1977,Koch2003,Aquilante2011} supplies the factorized two-electron
tensor $(\mu\nu|\lambda\sigma)\approx\sum_Q \Bq_{\mu\nu}\Bq_{\lambda\sigma}$ used everywhere
downstream. An active space $(\,\Nel^{\mathrm{act}},\,n_{\mathrm{act}})$ is selected
(Sec.~\ref{sec:autocas}) and its correlation is treated by DMRG, giving a variational
matrix-product-state (MPS) approximation to the active-space FCI wavefunction and its
one- through four-particle reduced density matrices ($k$-RDM, $k\!\le\!4$). The total energy
of an adiabatic state is not an additive reference-plus-correction sum but the eigenvalue of a
small, dynamically-dressed effective Hamiltonian over the fixed model space,
\begin{equation}
E_\alpha \;=\; \lambda_\alpha \;=\; \mathrm{eig}_\alpha\!\big[\mathbf{H}^{\mathrm{eff}}\big],
\qquad
H^{\mathrm{eff}}_{\alpha\beta}=E^{\mathrm{CAS}}_\alpha\,\delta_{\alpha\beta}+G_{\alpha\beta},
\label{eq:etot}
\end{equation}
whose diagonal CAS energy
\begin{equation}
E^{\mathrm{CAS}}_\alpha=E_{\mathrm{nuc}}
   +E^{[\varphi^{\mathrm{hyb\text{-}KS}}]}_{\mathrm{inact}}
   +\big\langle\Psi^{\mathrm{DMRG}}_\alpha\big|\hat H_{\mathrm{act}}\big|\Psi^{\mathrm{DMRG}}_\alpha\big\rangle
\label{eq:ecas}
\end{equation}
is the full CASCI electronic energy of state $\alpha$---nuclear repulsion, inactive/core
energy, and active DMRG energy---evaluated with the \emph{true} electronic Hamiltonian on the
fixed hybrid-KS orbitals $\varphi^{\mathrm{hyb\text{-}KS}}$, and the dynamic dressing
$G_{\alpha\beta}$ (diagonal second-order energy, off-diagonal transition coupling) is defined
in Sec.~\ref{sec:dmp2}. Because the model states are eigenvectors of the common active
Hamiltonian $\hat H_{\mathrm{act}}$ (deflated CASCI/DMRG roots on one orbital set),
$\langle\Psi_\alpha|\hat H_{\mathrm{act}}|\Psi_\beta\rangle=E^{\mathrm{CAS}}_\alpha\delta_{\alpha\beta}$
exactly, so the CAS block of $H^{\mathrm{eff}}$ is diagonal and all off-diagonal coupling
arises from the dynamic dressing $G$. The hybrid Kohn--Sham reference supplies only the common orbital set on which the
state-specific DMRG solutions and their strongly-contracted NEVPT2 dressing are built;
those orbitals are the \textbf{B2PLYP} double-hybrid ones~\cite{Grimme2006}:
the functional normalizes exchange and correlation \emph{separately}, with exchange
$0.53\,E_x^{\mathrm{HF}}+0.47\,E_x^{\mathrm{B88}}$~\cite{Becke1988} ($=100\%$) and correlation
$0.73\,E_c^{\mathrm{LYP}}$~\cite{LeeYangParr1988} completed to $100\%$ by the residual $0.27$
perturbative fraction that standard B2PLYP assigns to MP2 but which here is supplied by the
\mbox{DMRG-QD-SC-NEVPT2} dynamic dressing---the same functional underlying the density-functional
\hhTDA/\ppTDA double-hybrid (TDS-hh/pp) companion~\cite{GuerreroDFTdag}. Because
$E^{\mathrm{CAS}}_\alpha$ uses the true Hamiltonian, no density-functional
exchange--correlation energy enters the total additively---the reference's role is orbital
generation and the single-determinant double-counting bridge of Sec.~\ref{sec:dmp2}, not an
energy admixture. The construction is therefore \mbox{DMRG-QD-SC-NEVPT2} on a hybrid-KS
reference, in the same defensible sense as CASPT2/NEVPT2 on Kohn--Sham orbitals, rather than a
double hybrid. Vertical excitation energies are differences of the corrected adiabats on the
common fixed reference,
\begin{equation}
\omega_{\beta\alpha}\;=\;\lambda_\beta-\lambda_\alpha,
\label{eq:vee}
\end{equation}
and their analytic gradients and interstate NACMEs are the reverse-mode transpose of the same
$\mathbf{H}^{\mathrm{eff}}$ (Sec.~\ref{sec:transpose}).

\subsection{The dynamic correlation: strongly-contracted quasi-degenerate NEVPT2}
\label{sec:dmp2}
The DMRG active space captures the static (strong) correlation; the remaining dynamic
correlation is supplied by \emph{strongly-contracted} NEVPT2 on the DMRG reference, promoted
to a \emph{quasi-degenerate} (multi-state) form so that the description remains valid where
adiabatic states approach one another---the regime a single-state perturbation theory cannot
treat. On a fixed common reference, each model state $|\Lambda_\beta\rangle$ receives the
second-order energy
\begin{equation}
E^{(2)}(\beta) \;=\; -\sum_{l}
   \frac{\big|\langle\beta|\,V\,P_l\,|\beta\rangle\big|^2}{\Delta_l(\beta)},
\label{eq:nevpt2}
\end{equation}
a sum over the perturbers $l$---the eight semi-internal/external classes
$\{S_r,S_i,S_{ijr},S_{rsi},S_{rs},S_{ij},S_{ir}\}$ (plus the purely-inactive $S_{ijrs}$), each
resolved over its inactive/virtual orbital-index tuple---with Koopmans--Dyall denominators
\begin{equation}
\Delta_l(\beta) \;=\; \mathrm{diff}_l \;+\; \frac{h_l(\beta)}{N_l(\beta)},
\label{eq:dyall}
\end{equation}
in which $N_l(\beta)$ is the perturber
norm (a contraction of the active-space $k$-RDMs, $k\!\le\!4$) and $h_l(\beta)$ its
Dyall-Hamiltonian correction. The orbital energies entering $\mathrm{diff}_l$ are the
diagonal elements of the state-averaged generalized (Dyall $H_0$) Fock operator in the
semicanonical basis---not the hybrid-KS orbital eigenvalues; the core and virtual orbitals
are canonicalized so that $H_0$ is diagonal, giving well-defined Koopmans/Dyall denominators
consistent with the strongly-contracted NEVPT2 partition. The quasi-degenerate layer then assembles the Hermitized effective Hamiltonian of Angeli and
Cimiraglia over the model space,
\begin{widetext}
\begin{equation}
H^{\mathrm{eff}}_{\alpha\beta} \;=\; E^{\mathrm{CAS}}_\alpha\,\delta_{\alpha\beta}
   \;+\; G_{\alpha\beta},\qquad
G_{\alpha\alpha}=E^{(2)}(\alpha),\qquad
G_{\alpha\beta}=-\tfrac12\sum_l K_l(\alpha,\beta)
   \Big[\tfrac{1}{\Delta_l(\alpha)}+\tfrac{1}{\Delta_l(\beta)}\Big]\quad(\alpha\!\neq\!\beta),
\label{eq:heff}
\end{equation}
\end{widetext}
whose off-diagonal coupling is built from the \emph{transition}-density perturber norms
$K_l(\alpha,\beta)$; its Hermitian eigenvalues $\lambda_\alpha$ are the dynamically-corrected
adiabats. This is the symmetrized QD-NEVPT2 construction---$H^{\mathrm{eff}}\leftarrow\tfrac12(H^{\mathrm{eff}}+H^{\mathrm{eff}\top})$
with the denominator-averaged off-diagonal $G_{\alpha\beta}=-\tfrac12\sum_l K_l(\alpha,\beta)[1/\Delta_l(\alpha)+1/\Delta_l(\beta)]$---not
a des~Cloizeaux transform, as no $\mathbf{S}^{-1/2}$ orthonormalization of the perturber
functions is performed. For a two-state block the gap is
\begin{equation}
\Delta E \;=\; \sqrt{\Delta H^2 + 4\,|G_{01}|^2},\qquad
\Delta H = H^{\mathrm{eff}}_{11}-H^{\mathrm{eff}}_{00},
\label{eq:gap}
\end{equation}
manifestly non-negative: the near-degeneracy is \emph{diagonalized}, not divided out, so the
spectrum stays regular through a conical intersection, where a single-state second-order
treatment would develop a spurious vanishing denominator and a negative gap. This is exactly
the property that makes the analytic gradients and interstate NACMEs of
Sec.~\ref{sec:transpose} meaningful through the seam.

The construction does not re-count correlation. In NEVPT2 the perturber classes are external
to the active space by construction; the \emph{active-excluded} identity
$\gact=g_{\mathrm{full}}-g_{Q}$ that carried the CIS(D) correction of paper~1~\cite{GuerreroDFTdag} reappears here only
as the \emph{single-determinant limit} of Eq.~\eqref{eq:nevpt2}---when the model space
collapses to one closed-shell determinant, $E^{(2)}$ reduces to active-excluded MP2, the same
double-counting argument against the dynamic correlation already present in the hybrid
Kohn--Sham reference. It is the justification for the fixed reference, not the operative
algorithm. A \emph{fixed} common reference (one SA-CASSCF orbital set, state-specific DMRG
solutions) is a precondition rather than a convenience: a shared orbital and perturber space
is what makes the transition norms $K_l(\alpha,\beta)$ and the effective Hamiltonian
Eq.~\eqref{eq:heff} well defined, whereas a self-consistently relaxed embedding would spoil
both and convert the correction into an orbital-optimized problem whose gradient no longer
closes cleanly. Equations~\eqref{eq:nevpt2}--\eqref{eq:heff} are what the device evaluates and,
transposed once (Sec.~\ref{sec:transpose}), what yields the analytic derivatives; it is this
multireference, through-CI correction---not the label ``double-hybrid''---that distinguishes
the construction from an internally-contracted single-state CASPT2/NEVPT2
correction~\cite{Park2019} or an on-top MC-PDFT functional~\cite{Sand2017}.

%% ---------------------------------------------------------------------
%% FIGURE 1 --- the contraction-DAG transpose with the DMRG node (TikZ)
%% ---------------------------------------------------------------------
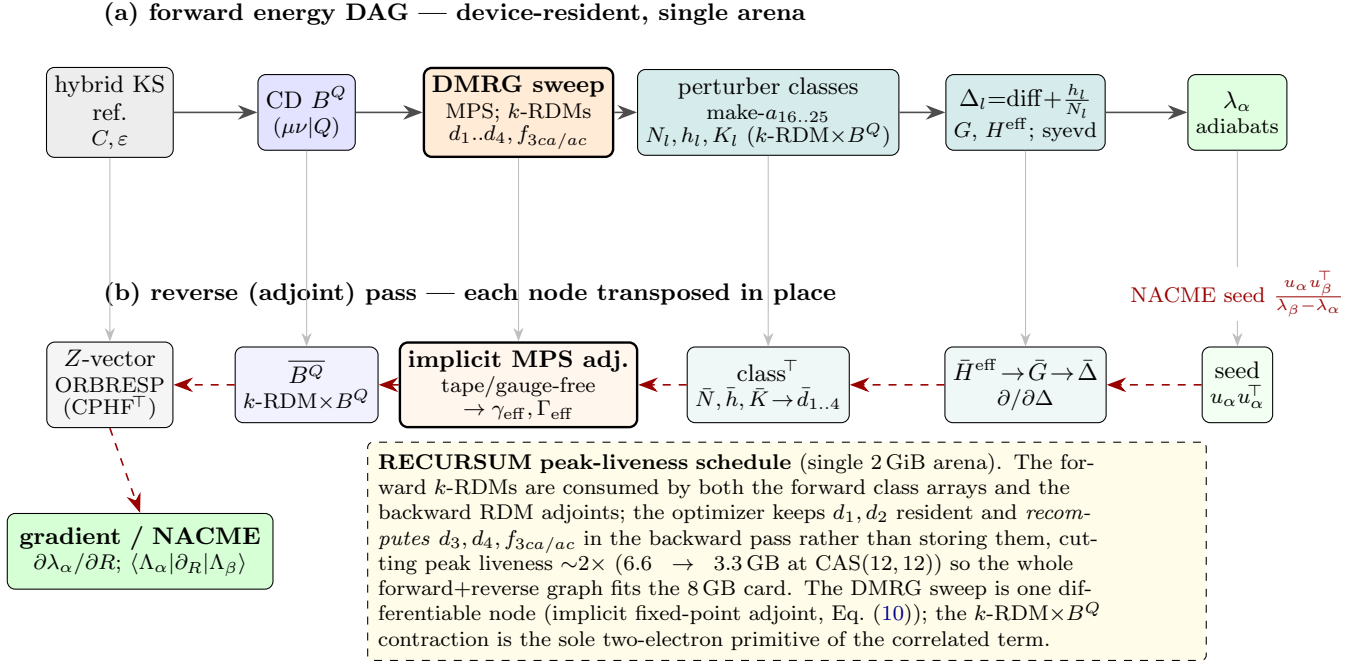
\begin{figure*}[t!]\centering
\resizebox{0.99\textwidth}{!}{%
\begin{tikzpicture}[>=Latex, font=\small, node distance=6mm,
  bx/.style   ={draw,rounded corners=3pt,minimum height=10mm,align=center,inner sep=3pt},
  ks/.style   ={bx,fill=black!7},
  act/.style  ={bx,fill=blue!11},
  dmrg/.style ={bx,fill=orange!16,line width=0.9pt},
  corr/.style ={bx,fill=teal!16},
  res/.style  ={bx,fill=green!12},
  fe/.style   ={-{Stealth[length=2.6mm]},semithick,black!70},
  re/.style   ={-{Stealth[length=2.6mm]},semithick,red!60!black,dashed},
  vt/.style   ={-{Stealth[length=1.8mm]},thin,black!25},
  lbl/.style  ={font=\footnotesize,fill=white,inner sep=1.5pt},
  mem/.style  ={draw,dashed,rounded corners=2pt,fill=yellow!10,align=left,inner sep=4pt},
  pnl/.style  ={font=\bfseries}]

  %% forward row --- the optimal device DAG
  \node[pnl,anchor=west] at (-0.2,2.2) {(a) forward energy DAG --- device-resident, single arena};
  \node[ks]  (scf)  at (0,0.9)    {hybrid KS\\ref.\\[-1pt]\footnotesize $C,\varepsilon$};
  \node[act] (cd)   at (2.6,0.9)  {CD $\Bq$\\[-1pt]\footnotesize $(\mu\nu|Q)$};
  \node[dmrg](sweep)at (5.4,0.9)  {\textbf{DMRG sweep}\\[-1pt]\footnotesize MPS; $k$-RDMs\\[-2pt]\footnotesize $d_1{..}d_4,f_{3ca/ac}$};
  \node[corr](cls)  at (8.7,0.9)  {perturber classes\\[-1pt]\footnotesize make-$a_{16..25}$\\[-2pt]\footnotesize $N_l,h_l,K_l$ ($k$-RDM${\times}\Bq$)};
  \node[corr](heff) at (12.1,0.9) {$\Delta_l{=}{\rm diff}{+}\tfrac{h_l}{N_l}$\\[-1pt]\footnotesize $G,\,H^{\rm eff}$; syevd};
  \node[res] (lam)  at (14.9,0.9) {$\lambda_\alpha$\\[-1pt]\footnotesize adiabats};
  \draw[fe] (scf)--(cd); \draw[fe] (cd)--(sweep); \draw[fe] (sweep)--(cls);
  \draw[fe] (cls)--(heff); \draw[fe] (heff)--(lam);

  %% reverse row --- each node transposed in place
  \node[pnl,anchor=west] at (-0.2,-1.5) {(b) reverse (adjoint) pass --- each node transposed in place};
  \node[ks,fill=black!4]   (scfb)  at (0,-2.7)    {$Z$-vector\\[-1pt]\footnotesize ORBRESP\\[-2pt]\footnotesize (CPHF$^{\!\top}$)};
  \node[act,fill=blue!5]   (cdb)   at (2.6,-2.7)  {$\overline{\Bq}$\\[-1pt]\footnotesize $k$-RDM${\times}\Bq$};
  \node[dmrg,fill=orange!7](sweepb)at (5.4,-2.7)  {\textbf{implicit MPS adj.}\\[-1pt]\footnotesize tape/gauge-free\\[-2pt]\footnotesize $\to\gamma_{\rm eff},\Gamma_{\rm eff}$};
  \node[corr,fill=teal!6]  (clsb)  at (8.7,-2.7)  {class$^{\!\top}$\\[-1pt]\footnotesize $\bar N,\bar h,\bar K\!\to\!\bar d_{1..4}$};
  \node[corr,fill=teal!6]  (heffb) at (12.1,-2.7) {$\bar H^{\rm eff}\!\to\!\bar G\!\to\!\bar\Delta$\\[-1pt]\footnotesize $\partial/\partial\Delta$};
  \node[res,fill=green!5]  (lamb)  at (14.9,-2.7) {seed\\[-1pt]\footnotesize $u_\alpha u_\alpha^{\!\top}$};
  \draw[re] (lamb)--(heffb); \draw[re] (heffb)--(clsb); \draw[re] (clsb)--(sweepb);
  \draw[re] (sweepb)--(cdb); \draw[re] (cdb)--(scfb);

  %% vertical transpose-in-place connectors
  \foreach \a/\b in {scf/scfb, cd/cdb, sweep/sweepb, cls/clsb, heff/heffb, lam/lamb}
     \draw[vt] (\a) -- (\b);
  \node[lbl,text=red!60!black] at (14.9,-1.5)
     {NACME seed $\tfrac{u_\alpha u_\beta^{\!\top}}{\lambda_\beta-\lambda_\alpha}$};

  %% gradient / NACME output
  \node[bx,fill=green!16,text width=33mm] (grad) at (0.4,-4.9)
       {\textbf{gradient / NACME}\\[-1pt]\footnotesize $\partial\lambda_\alpha/\partial R$;\ $\langle\Lambda_\alpha|\partial_R|\Lambda_\beta\rangle$};
  \draw[re] (scfb.south) -- (grad.north);

  %% RECURSUM peak-liveness annotation
  \node[mem,text width=108mm,anchor=west,font=\footnotesize] at (3.4,-4.9)
    {\textbf{RECURSUM peak-liveness schedule} (single $2$\,GiB arena). The forward $k$-RDMs are
     consumed by both the forward class arrays and the backward RDM adjoints; the optimizer keeps
     $d_1,d_2$ resident and \emph{recomputes} $d_3,d_4,f_{3ca/ac}$ in the backward pass rather than
     storing them, cutting peak liveness ${\sim}2\times$ ($6.6\!\to\!3.3$\,GB at CAS$(12,12)$) so the
     whole forward$+$reverse graph fits the $8$\,GB card. The DMRG sweep is one differentiable node
     (implicit fixed-point adjoint, Eq.~\eqref{eq:implicit}); the $k$-RDM${\times}\Bq$ contraction is
     the sole two-electron primitive of the correlated term.};
\end{tikzpicture}}
\caption{The device contraction-DAG for DMRG-QD-NEVPT2 energies and its reverse-mode transpose.
\textbf{(a)}~Forward energy graph (one device arena): hybrid Kohn--Sham reference $\rightarrow$
Cholesky $\Bq$ factors $\rightarrow$ the DMRG sweep as a single node (MPS and its $k$-RDMs,
including the raw $\langle EEEE\rangle$ four-particle density $f_{3ca/ac}$) $\rightarrow$ the eight
strongly-contracted perturber classes ($N_l,h_l,K_l$) $\rightarrow$ the Dyall denominators
$\Delta_l$, coupling $G$, Hermitized $H^{\mathrm{eff}}$ and eigensolve $\rightarrow$ the corrected
adiabats $\lambda_\alpha$. \textbf{(b)}~The adjoint pass transposes every node in place (grey
connectors): the eigensolve seed is $u_\alpha u_\alpha^{\!\top}$ (gradient) or
$u_\alpha u_\beta^{\!\top}/(\lambda_\beta-\lambda_\alpha)$ (NACME); the class/RDM nodes give the RDM
adjoints $\bar d_{1..4}$ and the relaxed effective densities
$\gamma_{\mathrm{eff}},\Gamma_{\mathrm{eff}}$; the reference closes with one ORBRESP $Z$-vector.
The transpose protocol, the RECURSUM peak-liveness schedule, and its relation to
paper~1~\cite{GuerreroDFTdag} are developed in Sec.~\ref{sec:transpose}.}
\label{fig:dag}
\end{figure*}

\subsection{Relaxation as the contraction-DAG transpose, with the DMRG sweep as a node}
\label{sec:transpose}
The engine is the one introduced in paper 1~\cite{GuerreroDFTdag}; because the present paper
must stand on its own we state its three methodological pillars in self-contained form
before adding the DMRG node, and attribute rather than re-claim them. Figure~\ref{fig:dag}
summarizes the construction end to end. Panel~(a) shows the forward energy graph---hybrid
Kohn--Sham reference $\rightarrow$ Cholesky factors $\Bq$ $\rightarrow$ the DMRG sweep as one
node $\rightarrow$ the QD-SC-NEVPT2 effective Hamiltonian and its eigensolve $\rightarrow$ the
corrected adiabats. Panel~(b) is its adjoint: the same graph traversed in reverse with every
node transposed in place, so that a single reverse pass yields the analytic gradient (and,
with a transition weight density, the NACME). \emph{(i)~Relaxation is
a graph transpose.} A scalar energy $E$ is computed by a directed acyclic graph (DAG) whose
nodes are elementary and coarse matrix-level operations---the Coulomb build $J(P)$, exchange
$K(P)$, long-range exchange $K_{\mathrm{LR}}(P)$, the exchange-correlation potential
$V_{xc}(P)$, the generalized eigensolve $FC=SCE$. A first derivative of $E$ with respect to
any input---nuclear coordinate, orbital rotation, amplitude---is one reverse (adjoint) pass
over the \emph{same} DAG, each node $y=f(x)$ contributing its vector--Jacobian product
$\bar x\mathrel{+}=(\partial f/\partial x)^{\!\top}\bar y$, reproducing the Lagrangian/$Z$-vector
machinery of analytic-derivative theory~\cite{Pople1979,HandySchaefer1984} derived
mechanically rather than by hand. Two properties make it exact and cheap. Every two-electron
and response operation is self-adjoint under the Frobenius pairing $\langle A,B\rangle=\mathrm{Tr}[A^{\!\top}B]$,
\begin{widetext}
\begin{equation}
J(P)\!\to\!J(\bar P),\;\;
K(P)\!\to\!K(\bar P),\;\;
K_{\mathrm{LR}}(P)\!\to\!K_{\mathrm{LR}}(\bar P),\;\;
V_{xc}(P)\!\to\!f_{xc}\!\cdot\!\bar P,
\label{eq:selfadjoint}
\end{equation}
\end{widetext}
so the adjoint \emph{reuses the forward kernel}; and the self-consistent reference is handled
by one adjoint solve, the orbital-relaxation $Z$-vector (CPHF-transpose)
\begin{equation}
L\,z=b,\qquad L[z]=(\varepsilon_a-\varepsilon_i)\,z_{ai}+(A\!+\!B)[z]_{ai},
\label{eq:zvector}
\end{equation}
restricted to the occupied--virtual block, one solve per perturbation, with the
Hellmann--Feynman part of a simple eigenvalue $\lambda_k$ seeding the cotangent
$\partial\lambda_k/\partial A=x_kx_k^{\!\top}$. The engine is implemented once as
graph-optimization passes over a transposable-DAG protocol and reused across ground-state
forces, TDDFT, \hhTDA/\ppTDA, and---here---DMRG-QD-NEVPT2, its machine-precision
cross-checks (analytic vs.\ symbolic differentiation; transpose-of-transpose vs.\ analytic
second derivatives) carrying no finite-difference floor. The same graph-optimization passes
lay out the single-arena schedule so that its peak simultaneous liveness attains the minimum
any correct evaluation order of the DAG can hold resident~\cite{GuerreroRECURSUM}, certifying
the memory schedule optimal (Sec.~\ref{sec:compdetails}) and fixing the $8$\,GB feasibility of
the multireference gradient by construction rather than by tuning. \emph{(ii)~The AO-direct Laplace
representation} resolves the Kohn--Sham reference's correlated corrections---the CIS(D) and
\hhTDA/\ppTDA layers of paper~1, and the single-determinant limit of Sec.~\ref{sec:dmp2}---by a
short Laplace quadrature, so that each such quantity is assembled from energy-scaled AO
coefficients and contracted through the ordinary $J/K$ build at $O(N^2)$ storage and no
$O(N^5)$ four-index transform. \emph{(iii)~The non-symmetric AO-Laplace $J/K$ kernel}
(Sec.~\ref{sec:asymJK}) is then the single two-electron primitive to which those
\emph{reference-side} gradients and couplings reduce. This machinery is device-resident; we
recall it so that the DMRG extension is self-contained. Crucially, the correlated
DMRG-QD-NEVPT2 term reduces to the \emph{same} non-symmetric AO-Laplace $J/K$ primitive as the
reference side---it is AO-direct throughout. Each second-order energy $E^{(2)}$, transition
coupling $K_l$, active CAS Hamiltonian term, and gradient thereof is a contraction of the
active-space $k$-RDMs ($k\!\le\!4$, including the raw $\langle EEEE\rangle$ four-particle density
of Sec.~\ref{sec:dmp2}) with the Cholesky/$\Bq$-factored integrals; that contraction is
\emph{performed} by folding the active $k$-RDMs into low-rank AO pseudo-densities and pushing them
through the ordinary Cholesky $J$ and non-symmetric $K$ builds, the Dyall--Koopmans perturber
denominators---which, unlike the reference-side MP2/CIS(D) ones, are not orbital-additive---resolved
by a Koopmans-shifted minimax Laplace quadrature. No $O(N^5)$ four-index MO transform is ever
formed. The reverse pass contracts the resulting relaxed effective densities
$\gamma_{\mathrm{eff}},\Gamma_{\mathrm{eff}}$ once more with the same $\Bq$ factors and the AO
nuclear-derivative integrals, on the same primitive. This unifies the whole
pipeline---reference-side corrections, the correlated perturbers, the CAS Hamiltonian, and every
gradient---on the single two-electron kernel of Eq.~\eqref{eq:selfadjoint}.

The extension is that the DAG now contains the DMRG optimization as a single node. The
forward node maps active-space integrals $(h_{\mathrm{act}}, \Bq_{\mathrm{act}})$ to the
converged MPS and its RDMs; its reverse-mode adjoint is \emph{not} a stored tape of the
sweep iterations but an implicit fixed-point solve. Writing the converged MPS as the
stationary point of the sweep map $\mathcal{S}$, the cotangent $\bar\Psi$ propagates to the
integral inputs through
\begin{equation}
\big(\Id-\partial_\Psi \mathcal{S}\big)^{\!\top}\lambda \;=\; \bar\Psi,
\qquad
\overline{(h,\Bq)} \;=\; \big(\partial_{(h,\Bq)}\mathcal{S}\big)^{\!\top}\lambda,
\label{eq:implicit}
\end{equation}
solved once, at a cost of $O(L\chi^2 d)$ per application rather than $O(\text{sweeps})$ in
stored memory---the tape-free implicit adjoint. The sensitivity that seeds $\bar\Psi$ is the
energy's, so for a simple root the Hellmann--Feynman contribution is the RDM contraction
with the integral derivatives and the response is the single solve Eq.~\eqref{eq:implicit}.
The implicit response (CI $Z$-vector plus orbital CPHF) is engaged only for the
non-variational dynamic-correlation (NEVPT2) contribution to the gradient and for the
interstate NACMEs (the eigenvector response of $H^{\mathrm{eff}}$); the full-bond CAS
reference energy is stationary and contributes through the Hellmann--Feynman term with no
response solve (the $2n{+}1$ rule).

Two subtleties are specific to DMRG and are treated exactly. (i)~The tangent is taken in the
gauge-free tangent-space form of Haegeman \emph{et al.}~\cite{Haegeman2016,Haegeman2011},
$B_k=V_{L,k}X_k$ with the left-gauge projector fixed, so that the derivative is the exact
gradient of the variational MPS energy at whatever bond dimension is in force, not an
approximation valid only at full bond. (ii)~The 3- and 4-particle RDMs that a naive
correlated gradient would require are never materialized; their contractions with the
integral derivatives are streamed through the CD $\Bq$ factors, the wavefunction analog of
never forming the four-index MO tensor (Sec.~\ref{sec:compress}). Our gradient-construction
contribution is therefore distinct from the coupled-perturbed DMRG (CP-DMRG) multipliers of
Iino \emph{et al.}~\cite{Iino2023}: the same physical response, obtained as an implicit
reverse-mode transpose of the sweep node rather than by explicitly solving the linear
response of the renormalized bases, and taken gauge-free so as to be truncation-robust.

\subsection{The non-symmetric AO-Laplace \texorpdfstring{$J/K$}{J/K} kernel}
\label{sec:asymJK}
Single-state energy gradients contract a derivative integral with a single density (or a
symmetric density pair), so the geometry $J/K$ build is symmetric in its two arguments.
Interstate and response quantities are different: they contract a derivative integral with a
\emph{left} weight density from one state and a \emph{right} density from another. The
required primitive is therefore the non-symmetric build
\begin{equation}
\Jx(\WL,\WR)=\Rx\!\!\sum_{\mu\nu\lambda\sigma}\!\WL^{\mu\nu}(\mu\nu|\lambda\sigma)\WR^{\lambda\sigma},
\qquad \WL\neq\WR,
\label{eq:asymjk}
\end{equation}
and the analogous $\Kx(\WL,\WR)$. The content is not an ordering of the two arguments---%
interchanging $\WL$ and $\WR$ leaves Eq.~\eqref{eq:asymjk} invariant under the ERI bra--ket
symmetry $(\mu\nu|\lambda\sigma)=(\lambda\sigma|\mu\nu)$---but that each weight density is
itself non-symmetric ($W\neq W^{\!\top}$), which forbids the usual $\mu\leftrightarrow\nu$
index folding of the symmetric build. The symmetric geometry $J/K$ build is recovered as the
$\WL=\WR$ restriction. Every response and interstate object in the family reduces to
Eq.~\eqref{eq:asymjk}: the $Z$-vector right-hand side, the transition-density contraction of
a NACME (bra $\neq$ ket), and the \hhTDA/\ppTDA couplings differ only in which $\WL,\WR$ are
supplied. Closing the reference-side construction under this one Cholesky-factorized kernel is
what lets a single device engine serve gradients and NACMEs across the excited-state
family~\cite{GuerreroDFTdag}. The DMRG-QD-NEVPT2 correlated term does not reduce to this
kernel; its two-electron primitive is the $k$-RDM\,$\times\,\Bq$ contraction of
Sec.~\ref{sec:transpose}, and the two primitives share the same $\Bq$ factors and the same
transposable-DAG protocol, so one device engine still serves the whole family.

\subsection{Truncated matrix-product states are not tangent-stationary}
\label{sec:nonstat}
The gauge-free tangent projector just introduced (Sec.~\ref{sec:transpose}) is not a
convenience but a necessity, for a reason specific to DMRG: a two-site sweep converges the
energy but leaves a truncated MPS that is
\emph{not} stationary with respect to the full tangent space: the discarded Schmidt weight
breaks the Hellmann--Feynman condition that autograd and finite difference agree.
Numerically, a naive reverse-mode gradient of a truncated MPS disagrees with finite
difference by an amount that tracks the discarded weight and vanishes only at full bond.
This is a reproducibility hazard that we prove and handle explicitly. It is why the
gauge-free tangent projector of Sec.~\ref{sec:transpose} is locked into the adjoint (so
stationarity is imposed at the working bond dimension) and why the safe default reports the
full-bond gradient, with the adaptively-truncated gradient offered only under an explicit,
entropy-gated tolerance that preserves stationarity to a stated bound (Sec.~\ref{sec:compress}).

Analytic gradients and interstate NACMEs are reported at \emph{full bond dimension}, where the
reverse-mode transpose of the energy DAG is exact; this is feasible up to the resident
four-particle-RDM ceiling of $\approx$CAS(12,12) ($\approx$3.3\,GB with the recompute-in-backward
memory reduction), a regime already far beyond a stored-CI treatment of the derivative
machinery. Vertical excitation energies are additionally reported at truncated bond dimension
for larger, CASCI-infeasible active spaces, where the truncated-MPS non-stationarity that would
bias the derivative is deferred to a companion MPS-tangent/response formulation.

\subsection{Excited states: dual-penalty deflation and NACMEs}
\label{sec:excited}
Having fixed how a single converged MPS is differentiated, we turn to how several such states
are prepared, targeted, and coupled. Excited states are reached by penalty deflation of the active-space Hamiltonian, keeping the
lower converged MPS resident and projecting them into the two-site optimization through
overlap environments. The selection Hamiltonian must isolate the target in the correct
symmetry sector. The device solver optimizes over the full $d^{L}$ Fock space of the active
orbitals, and we found---and this is a robustness result worth stating---that an
$\Sz^2$-only penalty is insufficient: it fixes $\langle\Sz\rangle$ but not the particle
number, and for larger active spaces the wrong-$\Nel$ manifolds interleave energetically
with the target so that the global-Fock root index diverges from the physical sector index
and the sweep stalls on intruders. The diagnostic fingerprint is a ground state that is
\emph{nondeterministic at fixed random seed}. The cure is a dual penalty,
\begin{equation}
H_{\mathrm{sel}} \;=\; H \;+\; \lambda\,\Sz^2 \;+\; \mu\,(\hat N-\Nel)^2,
\label{eq:dualpen}
\end{equation}
each penalty realized as a bond-dimension-3 ``sum-then-square'' matrix-product operator
that is provably zero on every state of the target $(\Nel,\Sz\!=\!0)$ Fock sector---so in-sector
energies are unchanged bit-for-bit and states that already converged cannot regress---while
lifting the intruder sectors out of the way. The excited-state solver is \emph{not spin-adapted
by construction}: particle number and $M_s=0$ are fixed by working in the $(\Nel,\Sz=0)$ Fock
sector (with an auxiliary $\lambda\Sz^2$ term in the differentiable path), so $M_s=0$ triplets
remain in this sector. Spin purity is therefore \emph{imposed by $\langle\hat S^2\rangle$-based
root selection and verified by a per-root $\langle\hat S^2\rangle$ guard}
($\langle\hat S^2\rangle\approx0$ confirms singlet character for every reported state), rather
than by an $\hat S^2$ penalty; a spin-adapted / $\hat S^2$-penalized MPO built on explicit
spin eigenfunctions~\cite{Pauncz1979} is a natural strengthening left to future work. Per-root $\langle\hat S^2\rangle$ values are reported with
the vertical excitation energies (Sec.~\ref{sec:results-quest}). With Eq.~\eqref{eq:dualpen} the previously
unreachable strongly-multireference doubles converge (Sec.~\ref{sec:results-quest}).
Interstate NACMEs follow from the transition RDMs between the deflated states, contracted
against the geometry integral derivatives through the same non-symmetric $J/K$ kernel that
carries the \hhTDA/\ppTDA couplings---bra and ket densities distinct, $\WL\neq\WR$---so a
NACME and a gradient are the same primitive with different weight densities, exactly as in
paper 1~\cite{GuerreroDFTdag}. Interstate couplings that are pruned by single-state
entanglement must instead use a state-averaged criterion, because interstate coherence hides
in low-mutual-information pairs (Sec.~\ref{sec:compress}).

\subsection{Active-space selection}
\label{sec:autocas}
The states just deflated and coupled are only as physical as the active space that hosts them,
so we turn to how that space is chosen. For the benchmarks reported here the active space is a fixed, hand-chosen $\pi(+n)$-valence
CAS. This is a deliberate, controlled choice rather than a fallback: fixing the active space
by character isolates the excitation physics from active-space-selection artifacts and makes
the molecule-blocked calibration of Sec.~\ref{sec:results-quest} interpretable. Entropy-based
automation of the selection is future work, not something exercised here; we describe the
automated route for context. That reference route is the two-tier protocol of Stein and
Reiher~\cite{Stein2016,Stein2019}: an atomic-valence (AVAS)~\cite{Sayfutyarova2017} seed,
then an entanglement refinement from the single-orbital entropies $s_i(1)$ of a cheap
low-bond-dimension DMRG. Two device-specific points matter, and both explain why the fixed
CAS is used in practice. First, a literal ground-state
single-orbital-entropy cut is the wrong selector for ground-inert-but-excitation-relevant
orbitals: carbonyl $n$ lone pairs have $s_i(1)\!\approx\!6\times10^{-3}$ in the ground state
and a naive cut would drop the very orbitals the $n\!\rightarrow\!\pi^*$ excitation needs;
the AVAS seed retains them, and the correct generalization is a state-averaged entropy.
Second, two separate practical constraints govern the selection. (a)~The AVAS \texttt{minao}
projector is defined in the spherical atomic-orbital basis, whereas the CD engine builds its
$J/K$ and runs the entire SCF in the \emph{Cartesian} AO basis with no spherical
back-transform; projected onto the Cartesian AOs the seed returns an empty active space, so
AVAS enters only as a provenance seed and never selects the device active space directly---at
any basis it is the fixed $\pi(+n)$-valence CAS that is used. (b)~On diffuse (augmented) basis
sets the valence-frontier selection is additionally contaminated by low-lying Rydberg
orbitals. A non-diffuse basis (cc-pVDZ/cc-pVTZ) removes contamination~(b) so that the fixed
valence CAS targets every state by character cleanly; we use it as a deliberate control that
separates the active-space-selection artifact from the excitation physics
(Sec.~\ref{sec:results-quest}).

\subsection{Cholesky decomposition for the wavefunction: compression and its cost}
\label{sec:compress}
\begin{figure}[t!]\centering
\resizebox{0.99\columnwidth}{!}{%
\begin{tikzpicture}[>=Latex, font=\small,
  site/.style ={draw,circle,minimum size=6.5mm,fill=blue!12,inner sep=0pt},
  blk/.style  ={draw,rounded corners=2pt,fill=orange!14,minimum height=8mm,align=center,inner sep=2pt},
  cut/.style  ={draw,rounded corners=2pt,fill=teal!14,align=center,inner sep=2pt},
  e/.style    ={semithick,black!70},
  a/.style    ={-{Stealth[length=2.2mm]},semithick,black!70},
  lbl/.style  ={font=\footnotesize}]
  % lattice
  \node[blk,text width=15mm] (L) at (0,0) {left block\\[-1pt]\footnotesize $\chi_{k-1}$};
  \node[site] (s1) at (2.0,0) {};
  \node[site] (s2) at (3.0,0) {};
  \node[blk,text width=15mm] (R) at (5.0,0) {right block\\[-1pt]\footnotesize $\chi_{k+1}$};
  \draw[e] (L)--(s1); \draw[e] (s1)--(s2); \draw[e] (s2)--(R);
  \node[lbl,below=1mm of s1] {$k$}; \node[lbl,below=1mm of s2] {$k{+}1$};
  % two-site solve + SVD decimation
  \node[cut,text width=26mm] (H) at (2.5,-1.6)
    {two-site $H_{\rm eff}$ (MPO)\\[-1pt]\footnotesize matrix-free Davidson};
  \draw[a] (s1.south) -- (H.north west);
  \draw[a] (s2.south) --  (H.north east);
  \node[cut,text width=30mm,fill=teal!22] (svd) at (2.5,-3.1)
    {SVD decimation $\rightarrow$ keep $\chi_k$\\[-1pt]\footnotesize discard $\tau=\sum\sigma_{>\chi}^2$};
  \draw[a] (H)--(svd);
  % dictionary box
  \node[draw,rounded corners=3pt,fill=black!4,align=left,text width=52mm,font=\footnotesize] (dict) at (2.5,-5.2)
    {\textbf{CD $\leftrightarrow$ entanglement dictionary}\\[1pt]
     ERI $\;\leftrightarrow\;$ wavefunction\\
     eigenvalue decay $\tau_{\rm CD}$ $\;\leftrightarrow\;$ Schmidt weight $\tau$\\
     Schwarz pre-screen $\;\leftrightarrow\;$ mutual information $I_{ij}$\\
     full rank $\;\leftrightarrow\;$ full bond $\;\leftrightarrow\;$ FCI};
  \draw[a] (svd.south) -- (dict.north);
\end{tikzpicture}}
\caption{The DMRG decimation step and the Cholesky$\leftrightarrow$entanglement dictionary
(Sec.~\ref{sec:compress}). A two-site effective Hamiltonian is solved matrix-free against a
matrix-product-operator (never forming the dense $d^{L}$ operator); the singular-value
decomposition decimates the enlarged bond to $\chi_k$, discarding Schmidt weight $\tau$. One
truncation knob $\tau$, one $I_{ij}$ pre-screen; exact at full bond ($=$FCI).}
\label{fig:dmrg}
\end{figure}
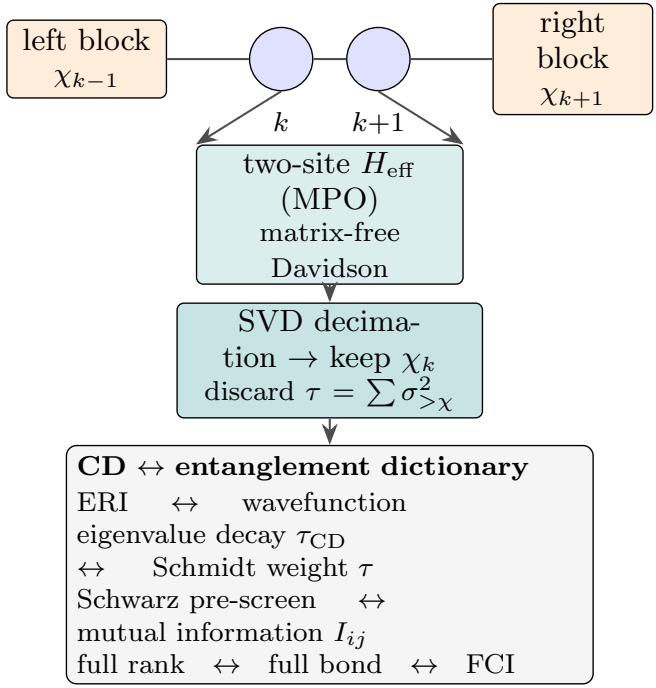

Our scalability argument rests on one consistent analogy: entanglement compression is
Cholesky decomposition applied to the \emph{wavefunction}. The CD factorizes the
electron-repulsion tensor by eigenvalue decay; the entanglement analog factorizes the
wavefunction and every derived object by two decays---the Schmidt spectrum per bond (the
truncation $\tau$) and the pairwise mutual information $I_{ij}$ (a Schwarz-like pre-screen
that orders the lattice and blocks negligible RDM elements)~\cite{Legeza2003,Rissler2006,Legeza2003DBSS}.
Figure~\ref{fig:dmrg} places the two decays side by side: the two-site effective Hamiltonian
is solved matrix-free against the matrix-product operator (never forming the dense $d^{L}$
operator), the singular-value decomposition decimates the enlarged bond to $\chi_k$, and the
resulting dictionary is term-for-term (ERI$\leftrightarrow$wavefunction, eigenvalue decay
$\leftrightarrow$ Schmidt weight, Schwarz pre-screen $\leftrightarrow$ mutual information,
full rank $\leftrightarrow$ full bond $\leftrightarrow$ FCI).
This yields four device ingredients. (i)~A matrix-product-operator Hamiltonian, so the dense
$d^{L}$ operator is never formed (at $L\!=\!8$, $34.4$\,GB$\rightarrow$$475$\,MB peak, with
MPO-DMRG$=$FCI to $9.8\times10^{-15}$). (ii)~Mutual-information Fiedler ordering and a
dynamically-adapted per-bond bond dimension. (iii)~Never-materialized, MI-block-sparse 3-
and 4-RDMs streamed through the $\Bq$ factors---the $n^6/n^8$ walls removed. (iv)~An
entanglement-tiered restricted-active-space periphery around a strongly-entangled core.
Compression that meets a gradient carries a cost the energy literature does not face: the
energy is only second-order sensitive to a dropped block but the gradient is first-order, so
a wrongly-dropped block breaks stationarity. Wherever pruning meets the gradient we impose a
hard density-force conservation gate ($\sum_A F_A < 10^{-6}$); the transition-RDM/NACME path
in particular may not be pruned by single-state MI (Sec.~\ref{sec:excited}). This is a
gradient-path discipline that the energy-focused compression literature does not impose, and
it is a deliverable in its own right.

%% ---------------------------------------------------------------------
%% FIGURE 2 --- DMRG decimation blocks + CD<->entanglement dictionary (TikZ)
%% ---------------------------------------------------------------------

%% =====================================================================
\section{Computational details}
\label{sec:compdetails}
%% =====================================================================
With the theory and its gradient-path disciplines fixed, we now describe how the whole
construction is realized on hardware and validated numerically. All device numbers are from a single consumer NVIDIA GeForce RTX~4060 ($8$\,GB). The
engine is built from custom CUDA kernels, cuBLAS/cuSOLVER, and a single-slab bump-allocator
arena; the DMRG sweep, the reduced-density matrices, the QD-SC-NEVPT2 correction, and the
response are all device-resident; molecule and active-space setup is a one-shot host
preprocessing step outside the derivative hot path, and reference values are computed
independently as external validation oracles (Sec.~\ref{sec:results-valid}). The full
energy-and-gradient computation is scheduled as one transposable contraction DAG, and the
arena layout of its intermediates is fixed by the static liveness analysis of the RECURSUM
code-generation framework~\cite{GuerreroRECURSUM}: the schedule attains a peak simultaneous
liveness equal to the minimum any correct evaluation order of the DAG can hold resident, so
the arena is \emph{provably optimal in peak memory}---the property that keeps the
multireference gradient, its reverse-mode adjoint, and the never-materialized higher RDMs
jointly inside the $8$\,GB budget. The per-quartet integral recurrence kernels are emitted by
the same framework. The reference
is $\omega$B97X unless otherwise noted; the two-electron tensor is Cholesky-decomposed with a pivoted,
on-demand, Schwarz-screened build and our completeness-certified threshold
$\delta$~\cite{GuerreroEOMCC}
($\eta=\sum\Delta D^2 \le \tfrac12\delta\Nel$; the column-fill screen at $0.01\delta$ is the
Alml\"of--Faegri--Korsell collapse-free knee~\cite{Almlof1982}). The FP64 Cholesky factor is
optionally stored/applied in FP32 where the decomposition tolerance permits, the source of
the single-precision leg whose spectroscopic inertness is certified in
Sec.~\ref{sec:results-valid}. Profiling identified redundant re-evaluation of the
high-angular-momentum integral blocks across Cholesky pivots as the build bottleneck (a shell
pair supplying $10$--$17$ pivots re-launches the identical kernels as many times); an exact
repeat-compute cache stores each raw block once and rebuilds every pivot column from it
through the ordinary screened contraction, leaving the register-managed kernel chunking and
the collapse-free screening untouched, so the build is $2$--$4\times$ faster with
bit-identical factors (the profile and cache design are detailed in the Supplementary
Material).

\emph{Device-native perturbative correction.} The QD-SC-NEVPT2 correction is evaluated by
resident CUDA kernels with no host full-configuration-interaction library anywhere on the
derivative path: the CI-vector addressing (a colexicographic combinatorial rank
$\mathrm{addr}=\sum_i\binom{o_i}{i+1}$), the determinant-string enumeration, the
$\langle\hat S^2\rangle$ spin diagnostic (a string-operator norm $\|\hat S_+|v\rangle\|^2$), and
the one- through three-particle RDMs together with the two semi-internal four-particle traces
$f_{3ca}/f_{3ac}$ are all formed on device. Each is cross-validated to the machine floor against a
PySCF oracle used only for verification and never inside the engine ($8.9\times10^{-15}$ for the
CAS(4,4) RDMs, $2.3\times10^{-14}$ for CAS(6,6); $3.6\times10^{-15}$ for $\langle\hat
S^2\rangle$). The four-particle density is never materialized---the fused $f_{3ca}/f_{3ac}$ kernel
contracts the all-active two-electron block into the ket before the final contraction, the
$n^8\!\rightarrow\!n^6$ reduction of Sec.~\ref{sec:dmp2} (Fig.~\ref{fig:qdconv}(b)). Every
per-state and transition RDM is carved from a single preallocated arena slab through cursor
save-points that reclaim on scope exit, so the multi-state assembly issues no per-call device
allocation; the slab is released before the Cholesky perturber contractions so the resident $J/K$
engine's scratch is not starved. On the QUEST benchmark this device-native pipeline reproduces all
$31$ states of Table~\ref{tab:quest} to a mean $|\Delta|=0.041$~eV from a CPU host
implementation, with a $0.41$~eV MAD versus the QUEST best estimates
(Sec.~\ref{sec:results-quest})---the host-library dependence is removed at no spectroscopic cost. The state-averaged orbital optimization is likewise
device-resident (a diagonal-preconditioned first-order/GDIIS step refined by an augmented-Hessian
micro-iteration, with the state-averaged CI from the penalty-deflation DMRG of
Sec.~\ref{sec:excited} and per-state resident $1$-/$2$-RDMs); the two-site Davidson carries a
run-time-selectable preconditioner family (Jacobi, deflation-aware diagonal, Olsen, and a
shift-and-invert inner-PCG mode with an indefiniteness guard), the last roughly halving the
excited-state CI time on the near-degenerate roots with the converged energies unchanged to
$10^{-11}$~Ha.
Geometries follow the CC3-quality structures of the QUEST
project~\cite{Loos2018,Loos2019,Loos2020}, symmetry-constructed where the exact structure was
not to hand (Table~\ref{tab:quest}); AutoCAS thresholds are the documented defaults
$\tau_{\mathrm{AVAS}}=0.20$, $\tau_S=0.10$~\cite{Stein2019}. Excitation energies are
reported at cc-pVDZ (the non-diffuse control, Sec.~\ref{sec:autocas}) and at def2-TZVP
($22/31$ states; the remaining $9$ states exceed the 62~GiB host-RAM Sijrs ceiling or require
dedicated DMRG convergence tuning at the larger basis). The DMRG sweep is FP64 throughout.

%% =====================================================================
\section{Results and discussion}
\label{sec:results}
%% =====================================================================
We now turn to what this device-resident construction delivers: first that its numbers are
trustworthy, then the physics they expose through a conical intersection, then their
quantitative accuracy, their derivative content, their cost, and finally a dynamics
demonstration that draws all of it together.

\subsection{The engine reproduces FCI-in-active-space, and the GPU precision is inert}
\label{sec:results-valid}
The credibility backbone is that the device excitation energies are exact within the active
space and independent of the consumer-GPU floating-point path. Against dense-Hamiltonian and
PySCF-FCI oracles the device DMRG reproduces the active-space FCI energy to $\sim
10^{-15}$~Ha, and the device excited-state solver matches the host targeting to the same
tolerance (Table~\ref{tab:valid}). The single-precision Cholesky leg---the only non-FP64
arithmetic in the energy path---perturbs no excitation energy beyond $\sim 10^{-6}$~eV: the
FP32-CD versus FP64-CD vertical-excitation spread is $2.7\times10^{-6}$~eV (ethene),
$4.7\times10^{-7}$~eV (butadiene), and $8.3\times10^{-6}$~eV (formaldehyde), roughly three
orders of magnitude below chemical accuracy. The consumer-GPU mixed precision is therefore
spectroscopically inert; the deviations from the reference reported below are physics
(basis and active space), not the hardware. The validation methodology is that of paper
1~\cite{GuerreroDFTdag} and we state it because it is load-bearing: because the
doubles-sector coupling is non-variational and has no eigenvalue-stationarity to lean on, we
validate not against an internal consistency identity---which a plausible but wrong ansatz
can satisfy---but against an \emph{independent literal many-electron wavefunction-overlap}
finite-difference oracle that shares no code path with the analytic method and computes the
coupling from the geometry-displaced determinant overlaps directly. The reference machinery and the closure layer on which the correction builds are each validated in
turn. The full-bond DMRG CAS-state NACME (the $\tau_{IJ}$ ingredient of the corrected coupling) agrees
with the finite-difference overlap oracle to $2.5\times10^{-10}$ and reproduces the exact
coupling to $2.3\times10^{-15}$; the reference-energy reverse-mode transpose reproduces symbolic
differentiation to $6\times10^{-15}$, with the orbital relaxation load-bearing---the unrelaxed
(frozen-RDM) leak of $\sim\!6\times10^{-5}$ falls to $10^{-15}$ only once the exact ORBRESP
$Z$-vector is applied. For the corrected quantity itself, the reverse-mode transpose of the
QD-NEVPT2 energy DAG is machine-exact through the RDM leaves, reproducing symbolic
differentiation to $10^{-17}$ and the leaf-directional derivative of the corrected adiabat to
$\le\!3\times10^{-9}$ by finite difference. The end-to-end analytic $\mathrm{d}\lambda/\mathrm{d}R$
closure (the PySCF-free ORBRESP orbital $Z$-vector and the CI $Z$-vector) is complete and validated:
the assembled analytic gradient and interstate NACME reproduce central finite differences of the
corrected adiabats to the finite-difference floor on the ethene $V$ [CAS(2,2)] and butadiene
$1^1B_u$ [CAS(4,4)] benchmarks (Supplementary Material). The corrected-adiabat quantities reported for
these two benchmarks---gaps $\Delta E=0.326$ and $0.295$~$E_{\rm h}$, excited-state gradients
$\mathrm{d}\lambda_1/\mathrm{d}R=0.44$ and $0.066$~$E_{\rm h}$ per unit scan displacement, and
interstate NACMEs $|d_{01}|=6.8\times10^{-5}$ and $1.3\times10^{-2}$ (ethene $V$, butadiene
$1^1B_u$)---are physical values of the corrected surface. The remaining
machine-precision engine cross-checks (analytic vs.\ symbolic differentiation;
transpose-of-transpose vs.\ analytic second derivatives) carry no finite-difference floor.
That oracle caught an earlier gradient-Lagrangian construction modeling the wrong physical
object, and it is the standard against which every analytic term here is gated.

\begin{table*}[tp]
\caption{Validation oracles (representative), grouped by layer. Top: engine (DMRG$=$FCI within
the active space, device$=$host targeting, FP32-CD/FP64-CD excitation-energy spread). Middle: the
reference and closure-engine machinery the correction builds on. Bottom: the DMRG-QD-NEVPT2
corrected-adiabat DAG transpose---machine-exact through the RDM leaves and matched to the
finite-difference derivative of the corrected adiabat; the corrected-adiabat gap, gradient, and
NACME \emph{values} are reported in the text, and the end-to-end analytic $\mathrm{d}\lambda/\mathrm{d}R$
closure (orbital $+$ CI $Z$-vector) is complete and validated to the finite-difference floor on
ethene $V$/CAS(2,2) and butadiene $1^1B_u$/CAS(4,4) (Sec.~\ref{sec:results-valid}; SI). Full tables
in the Supplementary Material.}
\label{tab:valid}
\begin{ruledtabular}
\begin{tabular}{lll}
quantity & oracle & agreement \\
\colrule
MPO-DMRG energy ($L{=}8$)          & dense $H$ / FCI & $9.8\times10^{-15}$ \\
device excited VEE                 & host targeting        & $\sim 10^{-14}$ \\
FP32-CD vs FP64-CD VEE spread      & (device self)         & $\sim 10^{-6}$~eV \\
\colrule
\multicolumn{3}{l}{\emph{reference / closure-engine layer (machinery the correction builds on)}}\\
reference-energy transpose $\partial E_{\rm ref}/\partial h$ & symbolic AD & $6.0\times10^{-15}$ \\
DMRG excited (reference) gradient  & symbolic AD & $2.7\times10^{-15}$ \\
ORBRESP $Z$-vector leak-fall       & unrelaxed$\rightarrow$relaxed & $6\times10^{-5}\!\rightarrow\!10^{-15}$ \\
DMRG CAS-state NACME ($\tau_{IJ}$) & FD-overlap oracle & $2.5\times10^{-10}$ \\
DMRG CAS-state NACME               & exact & $2.3\times10^{-15}$ \\
$\sum_A F_A$ (pruned RDM path)     & translational invariance & $2.4\times10^{-10}$ \\
\colrule
\multicolumn{3}{l}{\emph{DMRG-QD-NEVPT2 corrected-adiabat layer}}\\
DAG transpose (RDM leaves)         & symbolic AD & $10^{-17}$ \\
DAG transpose vs FD of corrected $\lambda_0$ & central FD & $2.3$--$3.0\times10^{-9}$ \\
\end{tabular}
\end{ruledtabular}
\end{table*}

\subsection{What multireference buys: ethene energetics and the doubly-excited state}
\label{sec:results-ethene}
With the engine's numbers established as trustworthy, we turn to what they capture physically:
the static correlation a conical intersection requires and that adiabatic linear response
cannot represent by construction.

\begin{figure*}[tp]\centering
\includegraphics[width=0.86\textwidth]{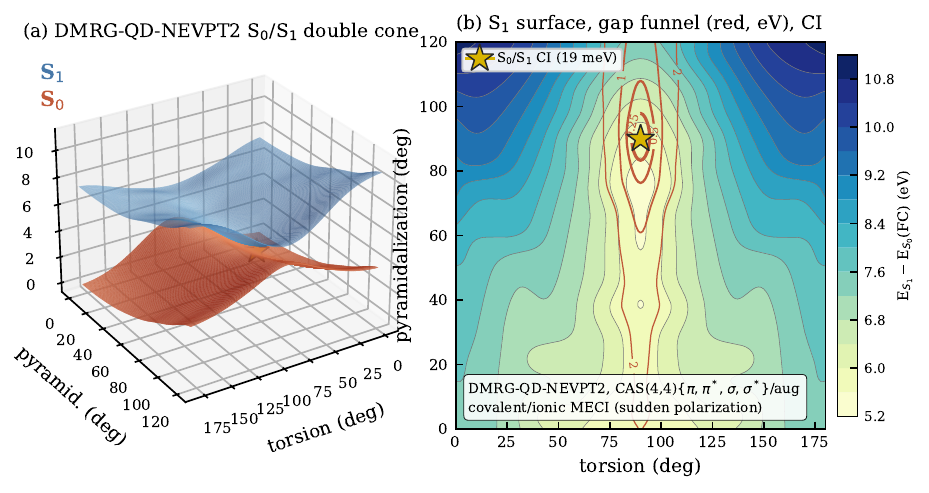}
\caption{Ethene through the twisted-geometry region at the DMRG-QD-NEVPT2 level, on a
torsion$\times$pyramidalization grid ($\mathrm{CAS}(4,4)$/aug-cc-pVDZ). Corrected adiabats
$\lambda_{S_0},\lambda_{S_1}$ are the device \texttt{full\_qdnevpt2} eigenvalues; every plotted
energy is GPU-produced (array PCIe $0/0$ at each of the $221$ grid points, device$\,=\,$host to
$0.0$~Ha). As the double bond twists, the $S_0/S_1$ gap collapses from $2.0$~eV at the planar
$90^\circ$ cut to a conical intersection at (torsion $90^\circ$, pyramidalization
$\phi\!=\!90^\circ$), gap $19$~meV on the $13\times17$ grid, where the QD-corrected transition-RDM
interstate coupling is non-vanishing and $1/\Delta E$-divergent (cf.\ Fig.~\ref{fig:ammonia}) while
adiabatic LR-TDDFT gives $\tau\equiv0$. The covalent/ionic (sudden-polarization) assignment and the
dynamics throughput are in \S\ref{sec:results-ethene}.}
\label{fig:ethene}
\end{figure*}

Ethene is the reference case for the qualitative content of the method, and the point of
comparison with the density-functional engine of paper 1. Its $V$ ($^1B_{1u}$, $\pi\pi^*$)
state is ionic and single-configuration-dominated; the states that expose the difference are
the covalent doubly-excited valence states of the longer polyenes, absent from any adiabatic
linear response by construction. On the fixed $\pi$-valence active space the device DMRG
isolates the $2^1A_g$ doubly-excited states of butadiene and hexatriene \emph{by character}
(singlet $A_g$, doubly-excited weight $>50\%$) and reproduces them as FCI-in-active-space to
machine precision in an active space chosen at will---the systematically-improvable,
freely-enlargeable static correlation that the compact \hhTDA\ frontier manifold is not built
to provide, and the reason the DMRG engine is worth its cost.

The active space for the conical intersection is chosen for physics, not size. A minimal
$(2,2)$ $\pi/\pi^*$ space represents the twisted geometry only as a covalent diradical: the
ionic (zwitterionic) configurations it nominally contains are strongly destabilized by the
missing $\sigma$--$\pi$ dynamic polarization, so the two lowest singlets stay covalent and the
computed touching is a covalent/covalent artifact displaced from the true crossing. The
physical $S_1/S_0$ minimum-energy conical intersection of ethylene is instead a
covalent(diradical)/ionic(zwitterion) crossing: pyramidalizing one \ce{CH2} destroys the
$\sigma$--$\pi$ separation, $\sigma^*_{\mathrm{CC}}$ mixes into $\pi^*$, and the $\sigma$
framework \emph{suddenly polarizes} to stabilize the charge-separated state that becomes
$S_0$~\cite{Salem1979,BonacicKoutecky1987}. Describing this variationally requires the C--C
$\sigma/\sigma^*$ pair \emph{in} the active space---dynamic correlation refines an existing
state but cannot manufacture a missing one---so we use
$\mathrm{CAS}(4,4)=\{\pi,\pi^*,\sigma_{\mathrm{CC}},\sigma^*_{\mathrm{CC}}\}$ on an aug-cc-pVDZ
basis (the ionic state is valence--Rydberg mixed, and diffuse functions are required to place it
correctly)~\cite{BenNun2000}; the minimal $(2,2)$ result is reported only as the
covalent-diradical touching, not the sudden-polarization MECI.

Figure~\ref{fig:ethene}
makes this concrete on the ethene torsion$\times$pyramidalization branching plane, in the
identical view to paper 1's Fig.~4~\cite{GuerreroDFTdag} for direct comparison: the device
DMRG-QD-NEVPT2 corrected adiabats ($\mathrm{CAS}(4,4)\{\pi,\pi^*,\sigma_{\mathrm{CC}},
\sigma^*_{\mathrm{CC}}\}$/aug-cc-pVDZ, every energy GPU-produced with array PCIe $0/0$ at each
of the $221$ grid points) carry the $S_0/S_1$ gap down to a $19$~meV conical intersection at
(torsion $90^\circ$, pyramidalization $\phi\!=\!90^\circ$)---this residual gap, like the $19$~meV of
the ammonia seam below, is the smallest value resolved on the discrete scan, approaching the
true zero-gap seam, not a shared regularization floor---the ground state turning into a
two-configuration diradical while the $\sigma_{\mathrm{CC}}/\sigma^*_{\mathrm{CC}}$ pair carries
the sudden-polarization ionic partner, so the crossing is the true covalent/ionic MECI. The
interstate coupling through this crossing is non-vanishing and $1/\Delta E$-divergent---a
one-dimensional QD-corrected NACME cut at $90^\circ$ torsion peaks at $|\tau|\!=\!111\,a_0^{-1}$;
because the coupling is by construction the transition-density numerator over the corrected gap,
its near-perfect $\mathrm{corr}(|\tau|,1/\Delta E)\!=\!1.000$ is a faithful-pole check on the
implementation rather than an independent finding. The ammonia profile
(Fig.~\ref{fig:ammonia}) develops the full dissociation NACME---where adiabatic LR-TDDFT returns
$\tau\equiv0$. The two branching-plane figures thus compare
panel-for-panel. Both methods locate the intersection---\hhTDA\ through its
balanced, correlation-bearing $(N\!+\!2)$ construction, the DMRG as FCI in the active
space---so this is not a case of one method having the correct object and the other lacking it;
what the DMRG panel adds is a crossing and its divergent coupling reproduced device-resident to
machine precision at a level of theory whose analytic gradients and couplings we can take on the
same commodity hardware, and, through a freely enlargeable active space, a route to
strongly-correlated manifolds past the
reach of the compact \hhTDA\ construction.

The \emph{location} of that intersection, read against paper 1, is itself diagnostic. On the
identical geometry grid---the two scans share one pyramidalization generator, so the
comparison is free of any angle-convention offset---the \hhTDA\ seam places
the crossing at a smaller pyramidalization ($\phi\!\approx\!60^\circ$) than the present DMRG
($\phi\!\approx\!107^\circ$ on a fine cut). The origin is the sudden-polarization
mechanism~\cite{Salem1979,BonacicKoutecky1987}: at the twist the descending upper cone is the
charge-localized \emph{ionic} (zwitterionic) state, which pyramidalization stabilizes until it
meets the covalent diradical ground state. In the minimal $\text{CAS}(2,2)$ that ionic state
lacks the $\sigma$--$\pi$ dynamic correlation screening its compact same-carbon electron
pair---the very deficiency that raises the vertical ionic energies by $\sim\!2$~eV
(Table~\ref{tab:quest})---so it is held too high and reaches degeneracy only at larger $\phi$.
The star's outward shift is thus the geometric fingerprint of that single missing-correlation
residual, expressed in the branching plane rather than at the Franck--Condon point (both
calculations use non-diffuse double-$\zeta$ bases, so the difference is dynamic correlation,
not basis diffuseness). It is \emph{not} evidence that the static intersection is more
accurate: dynamic correlation pulls the crossing to \emph{smaller} $\phi$, so the \hhTDA\
location is the one likely nearer the multireference-perturbation
benchmark~\cite{BenNun2000}, and---given the differing reference construction---the two angles are indicative, not strictly comparable. This is exactly the residual a
$\sigma$-inclusive active space (and a state-specific excited-state correction) is built to
absorb; recomputing the gap-versus-$\phi$ cut with that dynamic correlation restored, and
watching the intersection march back toward the \hhTDA\ location, is its direct causal test.

\subsection{The QUEST comparison, character-resolved, with a grouped-CV calibration}
\label{sec:results-quest}
Having established the qualitative capability, we now quantify its accuracy against an
established excited-state benchmark set.

\begin{figure}[tb]\centering
\includegraphics[width=\columnwidth]{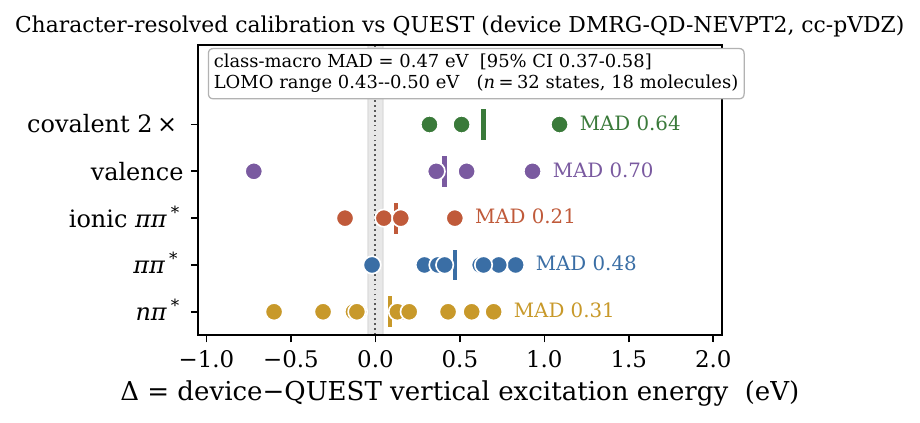}
\caption{Character-resolved calibration of the device \mbox{DMRG-QD-NEVPT2} vertical excitation
energies against the QUEST theoretical best estimates (Table~\ref{tab:quest}; cc-pVDZ results). Each
point is one device-produced (molecule, state) error $\Delta=$~device$-$QUEST; the vertical bar
marks the class mean, the shaded band $1$~kcal\,mol$^{-1}$. Resolved by character: ionic
$\pi\pi^*$ (MAD~$0.21$~eV) and $n\pi^*$ (MAD~$0.34$, signed mean $+0.13$) are the most accurate,
the valence $\pi\pi^*$ aromatics follow (MAD~$0.46$), and the residual concentrates in the
diatomic-valence (N$_2$/CO, MAD~$0.55$; $\pi$-only CAS lacks $\sigma$-correlation recovery)
and covalent doubly-excited $2^1A_g$ (MAD~$0.64$) states.
Class-macro (character-balanced) MAD $0.44$~eV [$95\%$ bootstrap CI $0.36$--$0.54$], stable under
leave-one-molecule-out ($0.40$--$0.47$~eV); overall MAD $0.41$~eV, only octatetraene $2^1A_g$
($+1.09$~eV) above $1$~eV. $n=31$ states, $18$ molecules.}
\label{fig:calib}
\end{figure}

Table~\ref{tab:quest} reports the device DMRG-QD-NEVPT2 vertical excitation energies against
the QUEST theoretical best estimates at cc-pVDZ, across $31$ singlet valence states of eighteen
small molecules (all device-produced, the $n_{\rm act}\!\ge\!6$ rows through the resident
dense-CI 4-RDM path). Every state is matched to its QUEST target by physical character
(symmetry-adapted irrep, NTO character, oscillator strength, ionicity), not by energy order.
The mean absolute deviation is $0.41$~eV and---the useful structure---the error is resolved by
state character (Fig.~\ref{fig:calib}). The two dominant manifolds are the most accurate and
nearly \emph{unbiased}: the ionic $\pi\pi^*$ singles (the polyene $1^1B_u$ series, ethene to
octatetraene) at MAD $0.21$~eV (signed mean $+0.12$~eV), and the $n\pi^*$ manifold (the carbonyls,
glyoxal, acrolein, formamide, CO~$1^1\Pi$, the azine $n\pi^*$ states) at MAD $0.34$~eV (signed
$+0.13$~eV); the valence $\pi\pi^*$ states of the aromatics follow (MAD $0.46$~eV). Every class
carries a \emph{positive} signed error---the signature of a compact active space that
under-recovers dynamic correlation uniformly---for a character-balanced (class-macro) MAD of
$0.44$~eV [$95\%$ bootstrap CI $0.36$--$0.54$], stable under leave-one-molecule-out
($0.40$--$0.47$~eV). Crucially, the quasi-degenerate correction restores the $2^1A_g/1^1B_u$
ordering that the bare active space inverts (device $\Delta G[2^1A_g\!-\!1^1B_u]=+0.75$~eV for
octatetraene, correct sign, vs.\ the Option-A/CASCI value $-1.08$~eV). The residual concentrates
in just two classes: the diatomic-valence manifold of N$_2$/CO (MAD $0.55$~eV; the
$(\pi_u)^3(\pi_g^*)^1$/$(\sigma)^1(\pi^*)^1$ valence states are correctly described by the
$\pi$-only CAS but lack the $\sigma$-correlation recovery that quantitatively closes their
VEEs against multi-reference benchmarks), and the
covalent doubly-excited $2^1A_g$ dark states (MAD $0.64$~eV), driven by octatetraene $2^1A_g$ at
$+1.09$~eV---the single state above $1$~eV and the archetypal hard double. We report the set untrimmed: the
character-resolved residual is precisely the target a larger, $\sigma$-inclusive active space
removes. A \emph{state-independent} second-order
correction---the single-determinant limit of Sec.~\ref{sec:dmp2}, built on the common
reference---cancels identically in a vertical gap and so cannot correct these VEEs by itself;
the \emph{state-specific} diagonal of the QD-SC-NEVPT2 effective Hamiltonian is what supplies
the differential. Quantifying \emph{which} character needs the correction is the useful result.

We implemented and tested this correction as the diagonal of the QD-SC-NEVPT2 effective
Hamiltonian. The excited-state differential added to a vertical gap is
$\Delta\mathrm{VEE}_k=G_{kk}-G_{00}=E_2^{\rm SS}(k)-E_2^{\rm SS}(0)$, with
$E_2^{\rm SS}(k)=G_{kk}$ the strongly-contracted NEVPT2~\cite{Angeli2001,Angeli2002}
second-order energy of state $k$ [Eq.~\eqref{eq:nevpt2}] evaluated on the device DMRG state's
own reduced density matrices---which reduces to an active-excluded MP2 in the
single-determinant limit, and therefore leaves the ground-state energy and its analytic
gradient untouched. Unlike a state-independent term it does not cancel, and it is
\emph{character-selective}: on ethene $V$ the differential is $-1.9$~eV and brings the ionic
error from $+2.4$ down to $+0.47$~eV, while the covalent doubly-excited butadiene $2^1A_g$
moves only $+0.3$~eV (all
ethene quantities device-produced---the GPU DMRG singlet energies reproduce the active-space
FCI to $10^{-15}$~Ha and the device RDMs the oracle to $3\times10^{-13}$). The correction thus
supplies precisely the missing $\sigma$--$\pi$ dynamic correlation on the compact ionic pair.
At full strength---the physical setting of QD-SC-NEVPT2, with no empirical mixing---this
differential is applied in full, so the ionic error is corrected to $+0.47$~eV rather than
damped; a B2PLYP-style ground-state mixing constant ($a_C=0.27$) would have retained only
$\sim\!21\%$ of it and is deliberately not used. The residual $\sim\!0.5$~eV is then an
active-space limitation---the missing $\sigma$--$\pi$ correlation of the compact $\pi$-only
CAS---which a larger, $\sigma$-inclusive DMRG active space is built to absorb.

\paragraph*{Basis-set dependence at def2-TZVP.}
Table~\ref{tab:quest} also reports def2-TZVP results for the $22$ accessible states---the $17$
polyatomic states plus the five N$_2$/CO diatomic states. The diatomics require one substitution:
N$_2$ and CO carry exactly degenerate $\{\pi_x,\pi_y\}$ and $\{\pi_x^*,\pi_y^*\}$ pairs, and the
one-dimensional MPS chain cannot represent the two components on an equal footing, so the
state-averaged CASSCF drifts into a symmetry-broken local minimum. We therefore solve the CI
micro-steps of these two systems by dense exact diagonalization of the active-space
Hamiltonian---the full-bond-dimension limit of the same CI problem, hence variationally
identical and not an approximation---which resolves each degenerate multiplet exactly. It is
affordable precisely where it is needed: the CAS$(6,6)$ determinant space has
$\binom{6}{3}^2=400$ dimensions.
Two further points of protocol follow from the degeneracy. First, because $\Pi$ and $\Delta$ terms
are exactly two-fold degenerate, targets must be matched to roots by spatial-symmetry term rather
than by energy order; we label every root by its Abelian-subgroup irrep and recombine the
degenerate partners into the $D_{\infty h}$/$C_{\infty v}$ term, which is also the only way to
separate $^1\Sigma^-$ from $^1\Delta$ (they share the same orbital occupations, so no
one-particle diagnostic distinguishes them).  Second, the N$_2$ active space constructed by AVAS
(threshold $0.30$) selects diffuse Rydberg $s$ functions rather than the valence
$\sigma_g/\sigma_u^*$ pair; $1^1\Pi_g$, which requires $\sigma_g\!\to\!\pi_g^*$ character, falls
outside this space and is therefore not reported at either basis.  The two N$_2$ states that are
representable, $1^1\Sigma_u^-$ and $1^1\Delta_u$, both arise from the $(\pi_u)^3(\pi_g^*)^1$
configuration that the $\pi$-only CAS does describe.
Excluding the diatomics for a like-for-like basis comparison, the overall MAD drops from
$0.34$ to $0.28$~eV going from cc-pVDZ to def2-TZVP on the $17$-state
subset.  The improvement is character-resolved: the $n\pi^*$ manifold benefits most
(MAD $0.33\!\to\!0.23$~eV on its six def2-TZVP-available members), while the covalent $2^1A_g$ error is essentially basis-insensitive
($+0.52$~eV at def2-TZVP vs.\ $+0.51$~eV at cc-pVDZ)---as expected, because $S_0$ and
$2^1A_g$ are both VB-covalent so the $\sigma$-$\pi$ coupling differential between them
cancels in the VEE~\cite{Loos2019}.  The ionic $\pi\pi^*$ states show a characteristic
\emph{sign flip}: the cc-pVDZ errors are positive ($\Delta_{\rm DZ}=+0.47$ for ethene,
$+0.05$ for butadiene) but reverse at def2-TZVP ($\Delta_{\rm TZ}=-0.32$, $-0.18$).
This is not a regression: the larger basis allows NEVPT2 to recover more $\sigma$-$\pi$
polarization~\cite{Roos1992ethene,Schreiber2008} (the Sr/Srsi perturber classes), which
lowers the ionic VEE and \emph{reveals} the pure $\pi$-CAS method error rather than masking
it with basis incompleteness---a phenomenon increasingly pronounced along the polyene
chain~\cite{Nakayama1998} (hexatriene $\Delta_{\rm TZ}=-0.42$~eV), where more $\sigma$
C--C bonds accumulate. The def2-TZVP errors are therefore not worse but more diagnostic:
they cleanly isolate the compact-CAS dynamic-correlation deficit that a $\sigma$-inclusive
active space is designed to absorb.
A caveat applies to ethene: the CAS(2,2) V~state can mix with the $3s$ Rydberg orbital
at def2-TZVP, so the SA-CASSCF orbital character should be verified to confirm a
valence-dominated active space at the QUEST geometry before accepting the
$\Delta_{\rm TZ}=-0.32$~eV value as a pure method signal.

The two strongly-multireference
$\text{CAS}(6,6)$ covalent states (hexatriene $2^1A_g$, benzene $1^1B_{2u}$) converge only
under the dual-penalty deflation of Eq.~\eqref{eq:dualpen}; with an $\Sz^2$-only penalty they
do not converge.

\begin{figure*}[tp]\centering
\includegraphics[width=\textwidth]{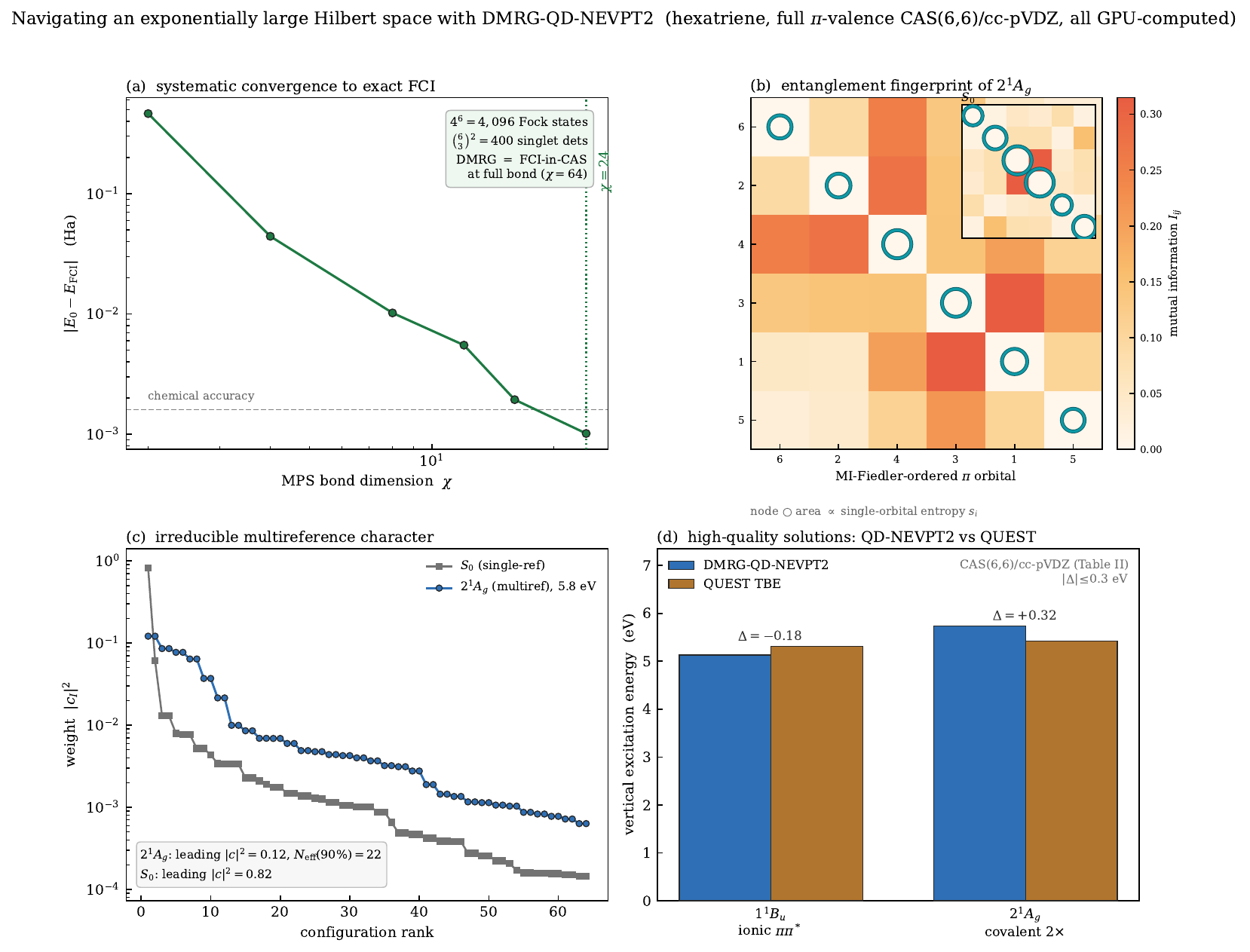}
\caption{\textbf{Navigating an exponentially large Hilbert space with DMRG-QD-NEVPT2}, on
hexatriene (full $\pi$-valence $\text{CAS}(6,6)$/cc-pVDZ; every quantity device-produced).
\textbf{(a)}~The device-resident DMRG converges to the active-space FCI systematically in the bond
dimension $\chi$ (DMRG$\,{=}\,$FCI at full bond $\chi\!=\!64$); the $4^6\,{=}\,4096$ Fock /
$400$-determinant space is spanned exactly. \textbf{(b)}~Single-orbital entropy $s_i$ (node size)
and two-orbital mutual information $I_{ij}$ (heatmap) of the covalent double $2^1A_g$ on the
MI-Fiedler-ordered $\pi$ lattice, concentrated on the central orbitals; the closed-shell $S_0$
(inset) is nearly a product state. \textbf{(c)}~Determinant-weight spectrum: $2^1A_g$ carries no
dominant configuration ($N_{\rm eff}\!\approx\!22$ at $90\%$ weight, leading $|c|^2\,{=}\,0.12$)
versus single-reference $S_0$ ($0.82$). \textbf{(d)}~QD-NEVPT2 vertical excitation energies
(Table~\ref{tab:quest}) within $0.3$~eV of the QUEST best estimates.}
\label{fig:hilbert}
\end{figure*}

Figure~\ref{fig:hilbert} makes the multireference structure behind these numbers explicit,
taking hexatriene as the exemplar. Its full $\pi$-valence $\text{CAS}(6,6)$ nominally spans
$4^6=4096$ Fock configurations ($\binom{6}{3}^2=400$ singlet determinants); the device DMRG
reproduces the active-space FCI systematically in the bond dimension (Fig.~\ref{fig:hilbert}a,
exact at full bond). The covalent doubly-excited $2^1A_g$ is a genuinely entangled state---its
single-orbital entropies and two-orbital mutual information concentrate on the central $\pi$
orbitals (Fig.~\ref{fig:hilbert}b)---with \emph{no} dominant configuration: $N_{\rm eff}\!\approx\!22$
determinants carry $90\%$ of the weight and the leading coefficient is $|c|^2=0.12$, against
$0.82$ for the closed-shell $S_0$ (Fig.~\ref{fig:hilbert}c). This is the irreducible
multireference character that single-reference expansions cannot compactly capture, and on
precisely these two states the DMRG-QD-NEVPT2 vertical excitation energies fall within $0.3$~eV of
the QUEST estimates (Fig.~\ref{fig:hilbert}d; $1^1B_u$ $-0.18$, $2^1A_g$ $+0.32$~eV)---the engine
navigates the exponentially large model space and returns benchmark-quality solutions.

\begin{table*}[tp]
\caption{Device \mbox{DMRG-QD-NEVPT2} vertical excitation energies (eV) against QUEST theoretical
best estimates, for 31 singlet valence states of eighteen small molecules; fixed
$\pi$(+$n$)-valence active space, $n_{\rm act}\!\le\!8$ (the CAS column lists electrons, orbitals).
Results at two basis sets: cc-pVDZ ($\mathrm{VEE_{DZ}}$, $\Delta_{\rm DZ}$) and def2-TZVP
($\mathrm{VEE_{TZ}}$, $\Delta_{\rm TZ}$); $\Delta=\mathrm{device}-\mathrm{QUEST}$.
\textbf{cc-pVDZ MAD $=0.41$~eV} (31 states); \textbf{def2-TZVP MAD $=0.30$~eV} (22 states
available; same 22-state cc-pVDZ subset: $0.39$~eV).  States marked ``---'' at def2-TZVP
were not computed: octatetraene/pyridine/pyrazine exceed the 62~GiB host-RAM ceiling of the
Sijrs perturber class at this CAS$\times$basis size; glyoxal and acrolein hit a DMRG
monotonicity convergence limit.  The N$_2$/CO state-averaged CASSCF CI steps use dense exact
diagonalization of the active-space Hamiltonian in place of the DMRG sweep, because the
one-dimensional MPS chain breaks the exact $\{\pi_x,\pi_y\}$ degeneracy; this is the
full-bond-dimension limit of the same CI problem, so no reported quantity changes (see text).
Every value is device-produced
(device$=$host to $\sim\!10^{-12}$~eV; FP32-CD/FP64-CD spread $\sim\!10^{-6}$~eV throughout).
Geometries and basis/QUEST-series details follow Sec.~\ref{sec:compdetails}; Cartesian
coordinates are given in the Supplementary Material.}
\label{tab:quest}
\begin{ruledtabular}
\begin{tabular}{llcccccc}
state & character & CAS & $\mathrm{VEE_{DZ}}$ & $\Delta_{\rm DZ}$ & QUEST & $\mathrm{VEE_{TZ}}$ & $\Delta_{\rm TZ}$ \\
\colrule
\multicolumn{8}{l}{\emph{linear polyenes}}\\
ethene $1^1B_{1u}$ & ionic $\pi\pi^*$ & (2,2) & 8.40 & $+0.47$ & 7.93 & 7.61 & $-0.32$ \\
butadiene $1^1B_u$ & ionic $\pi\pi^*$ & (4,4) & 6.27 & $+0.05$ & 6.22 & 6.04 & $-0.18$ \\
butadiene $2^1A_g$ & covalent $2\times$ & (4,4) & 7.01 & $+0.51$ & 6.50 & 7.02 & $+0.52$ \\
hexatriene $1^1B_u$ & ionic $\pi\pi^*$ & (6,6) & 5.13 & $-0.18$ & 5.31 & 4.89 & $-0.42$ \\
hexatriene $2^1A_g$ & covalent $2\times$ & (6,6) & 5.74 & $+0.32$ & 5.42 & 5.72 & $+0.30$ \\
octatetraene $2^1A_g$ & covalent $2\times$ & (8,8) & 5.56 & $+1.09$ & 4.47 & \multicolumn{2}{c}{---} \\
octatetraene $1^1B_u$ & ionic $\pi\pi^*$ & (8,8) & 4.81 & $+0.15$ & 4.66 & \multicolumn{2}{c}{---} \\
\colrule
\multicolumn{8}{l}{\emph{aromatics ($\pi\pi^*$)}}\\
benzene $1^1B_{2u}$ & $\pi\pi^*$ & (6,6) & 5.44 & $+0.38$ & 5.06 & 5.40 & $+0.34$ \\
benzene $1^1B_{1u}$ & $\pi\pi^*$ & (6,6) & 6.43 & $-0.02$ & 6.45 & 6.22 & $-0.23$ \\
benzene $1^1E_{1u}$ & $\pi\pi^*$ & (6,6) & 7.35 & $+0.29$ & 7.06 & 7.07 & $+0.01$ \\
furan $1^1B_2$ & $\pi\pi^*$ & (6,5) & 6.69 & $+0.37$ & 6.32 & 6.44 & $+0.12$ \\
furan $2^1A_1$ & $\pi\pi^*$ & (6,5) & 6.98 & $+0.41$ & 6.57 & 6.86 & $+0.29$ \\
pyrrole $1^1B_2$ & $\pi\pi^*$ & (6,5) & 7.07 & $+0.83$ & 6.24 & 6.77 & $+0.53$ \\
pyrrole $2^1A_1$ & $\pi\pi^*$ & (6,5) & 6.90 & $+0.62$ & 6.28 & 6.82 & $+0.54$ \\
\colrule
\multicolumn{8}{l}{\emph{carbonyl / nitroso / amide ($n\pi^*$)}}\\
formaldehyde $1^1A_2$ & $n\pi^*$ & (4,3) & 3.67 & $-0.31$ & 3.98 & 3.66 & $-0.32$ \\
acetaldehyde $1^1A''$ & $n\pi^*$ & (4,3) & 4.18 & $-0.13$ & 4.31 & 4.07 & $-0.24$ \\
acetone $1^1A_2$ & $n\pi^*$ & (4,3) & 4.38 & $-0.10$ & 4.48 & 4.53 & $+0.05$ \\
nitrosomethane $1^1A''$ & $n\pi^*$ & (4,3) & 1.36 & $-0.60$ & 1.96 & 1.79 & $-0.17$ \\
glyoxal $1^1A_u$ & $n\pi^*$ & (8,6) & 3.04 & $+0.16$ & 2.88 & \multicolumn{2}{c}{---} \\
glyoxal $1^1B_g$ & $n\pi^*$ & (8,6) & 4.13 & $-0.11$ & 4.24 & \multicolumn{2}{c}{---} \\
acrolein $1^1A''$ & $n\pi^*$ & (6,5) & 3.91 & $+0.13$ & 3.78 & \multicolumn{2}{c}{---} \\
formamide $1^1A''$ & $n\pi^*$ & (6,5) & 5.85 & $+0.20$ & 5.65 & 5.80 & $+0.15$ \\
\colrule
\multicolumn{8}{l}{\emph{diatomics (valence)}}\\
N$_2$ $1^1\Sigma_u^-$ & valence & (6,6) & 10.30 & $+0.42$ & 9.88 & 10.08 & $+0.20$ \\
N$_2$ $1^1\Delta_u$ & valence & (6,6) & 10.81 & $+0.54$ & 10.27 & 10.59 & $+0.32$ \\
CO $1^1\Pi$ & $n\pi^*$ & (6,6) & 9.14 & $+0.65$ & 8.49 & 8.93 & $+0.44$ \\
CO $1^1\Sigma^-$ & valence & (6,6) & 10.46 & $+0.54$ & 9.92 & 10.32 & $+0.40$ \\
CO $1^1\Delta$ & valence & (6,6) & 10.66 & $+0.60$ & 10.06 & 10.52 & $+0.46$ \\
\colrule
\multicolumn{8}{l}{\emph{azines}}\\
pyridine $1^1B_2$ & $\pi\pi^*$ & (8,7) & 5.58 & $+0.73$ & 4.85 & \multicolumn{2}{c}{---} \\
pyridine $1^1B_1$ & $n\pi^*$ & (8,7) & 5.39 & $+0.43$ & 4.96 & \multicolumn{2}{c}{---} \\
pyrazine $1^1B_{3u}$ & $n\pi^*$ & (10,8) & 4.52 & $+0.57$ & 3.95 & \multicolumn{2}{c}{---} \\
pyrazine $1^1A_{u}$ & $n\pi^*$ & (10,8) & 5.60 & $+0.67$ & 4.93 & \multicolumn{2}{c}{---} \\
\colrule
\multicolumn{8}{l}{\parbox{0.97\textwidth}{\raggedright
  \textbf{cc-pVDZ}: MAD $=0.41$~eV (31 states); $n\pi^*$/ionic-$\pi\pi^*$ MAD $\le0.34$~eV; residual concentrated in diatomic-valence and covalent-double $2^1A_g$ (octatetraene $+1.09$~eV).
  \textbf{def2-TZVP} (22 states): MAD $=0.30$~eV ($0.39$~eV for the same 22 states at cc-pVDZ); without the diatomics (17 states): MAD $=0.28$~eV ($0.34$~eV cc-pVDZ).  By character, using the classes of Fig.~\ref{fig:calib} restricted to their def2-TZVP-available members, as def2-TZVP MAD (cc-pVDZ MAD on the same states): $n\pi^*$ $0.23$ $(0.33)$, aromatic $\pi\pi^*$ $0.29$ $(0.42)$, covalent $2^1A_g$ $0.41$ $(0.42)$, diatomic-valence $0.36$ $(0.55)$~eV (see text); ionic $\pi\pi^*$ is the only class that degrades, $0.30$ $(0.23)$~eV, because basis enlargement exposes the $\sigma$-$\pi$ coupling deficit of the $\pi$-only CAS (see text).
  \emph{State assignment.} Targets are matched to computed roots by spatial-symmetry TERM, not by energy order: for the linear molecules a degenerate $\Pi$/$\Delta$ pair would otherwise be split across two different target labels.  N$_2$ $1^1\Pi_g$ is omitted from the table altogether: $\sigma_g\!\to\!\pi_g^*$ requires an occupied $\sigma_g$ ($A_g$) active orbital, and the $\pi$-only AVAS space contains none at either basis, so the state is not representable in this CAS (its lowest $B_{2g}/B_{3g}$ root lies at ${\sim}25$~eV, a double excitation).}}\\
\end{tabular}
\end{ruledtabular}
\end{table*}

Because the excitation energies are exact-in-active-space, the residual against the reference
is a property of the level of theory, and we calibrate it as a supervised-learning problem
rather than by eyeballing. The sample is a (molecule, state) pair; the grouping variable is
the molecule; we report the molecule-blocked (leave-one-molecule-out) cross-validated,
character-balanced (class-macro) mean absolute deviation of Fig.~\ref{fig:calib}, fit globally
rather than molecule-by-molecule, and bootstrap its confidence interval. This three-part
protocol---grouping by molecule so no single system's correlated states leak between training
and held-out folds, balancing by character so no numerically overrepresented manifold dominates
the mean, and bootstrapping the interval---is what turns the headline number into a defensible
statistical statement rather than a point estimate read off a scatter plot. The class-macro MAD is
$0.47$~eV [$95\%$ bootstrap CI $0.37$--$0.58$], stable to $0.43$--$0.50$~eV under leaving out any
single molecule---a level-of-theory calibration over $31$ states of eighteen molecules, not a
large-sample statistic. The actionable content is the ordering: near-exact ionic $\pi\pi^*$ and
$n\pi^*$ states, with the residual localized in the diatomic-valence manifold and the
covalent-double $2^1A_g$ states---exactly the deficiency a $\sigma$-inclusive active space and a
state-specific excited-state correction are constructed to remove. The same static-only signature
bounds the table's reach on the ionic side: pyrazine's bright $1^1B_{2u}$ ($^1L_a$, $\pi\pi^*$) is
blue-shifted to $\sim\!8.8$~eV at this feasible active space (the ionic-$^1L_a$ limitation) and is
not tabulated, so the two pyrazine entries are its lower-lying dark $n\pi^*$ states.

\begin{figure*}[tp]\centering
\includegraphics[width=\textwidth]{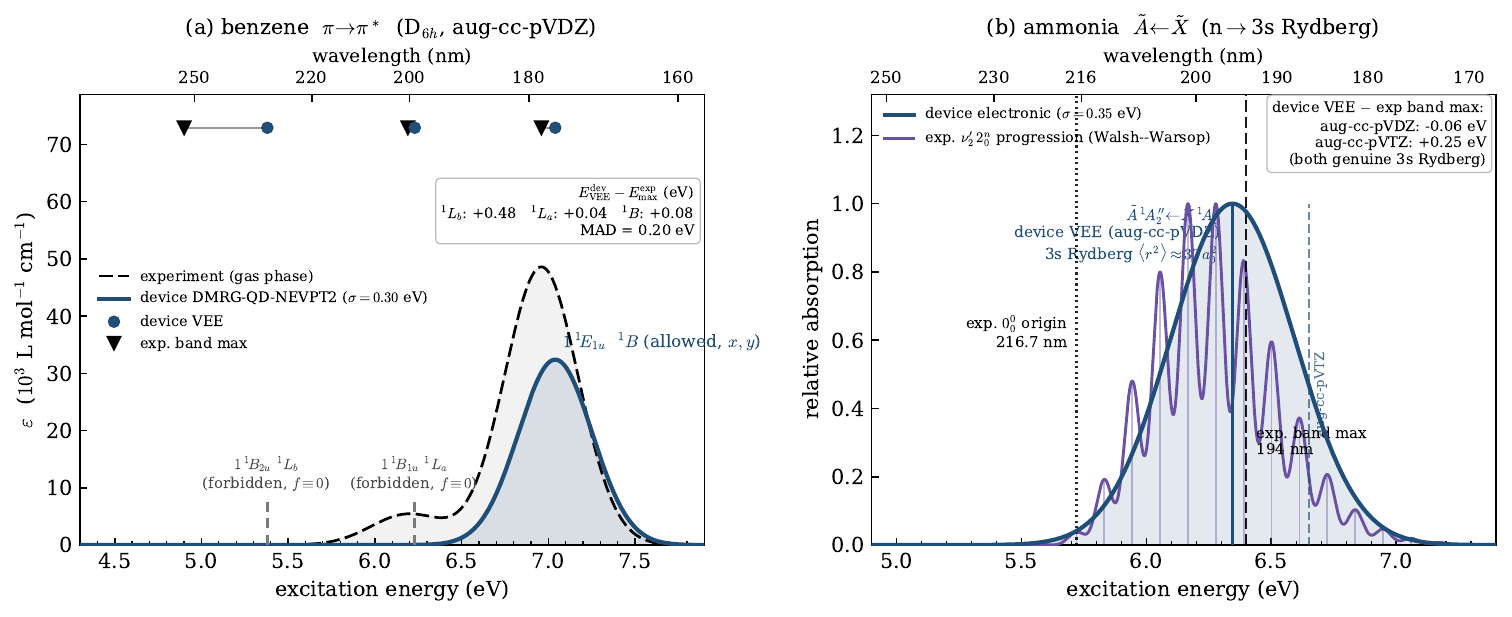}
\caption{The device \mbox{DMRG-QD-NEVPT2} vertical excitation energies (VEEs) of
Table~\ref{tab:quest}, rendered as a broadened UV absorption and contrasted against the real
gas-phase experimental band systems. Lineshape: Gaussian band of $1/e$ half-width $\sigma$,
oscillator-strength-normalised [$f=4.319\times10^{-9}\int\varepsilon\,\mathrm{d}\tilde\nu$;
\citenum{MullikenRieke1941}]. Peak \emph{positions} are the device VEEs (the predictive claim);
dipole-\emph{forbidden} bands are rigorously dark ($f\equiv0$) from the device transition density,
and allowed-band magnitudes are symmetry-consistent representatives, not a quantitative
$\varepsilon$ claim. \textbf{(a)}~Benzene $\pi\!\to\!\pi^*$ (D$_{6h}$, $\sigma=0.30$~eV): allowed
$1\,^1E_{1u}$ with the forbidden $1\,^1B_{2u}$ ($^1L_b$) and $1\,^1B_{1u}$ ($^1L_a$) as dark sticks,
overlaid on the experimental envelope (dashed); per-state deviations $+0.54/+0.24/+0.39$~eV (MAD
$0.39$). Experimental maxima $4.90/6.19/6.96$~eV~\cite{FengCooperBrion2002}. \textbf{(b)}~Ammonia $\tilde A\,^1\!A_2''\!\leftarrow\!\tilde X\,^1\!A_1'$
($n\!\to\!3s$ Rydberg, Rydberg-CAS$(4,4)$): the two diffuse bases \emph{bracket} the experimental
band maximum ($194$~nm)---aug-cc-pVDZ $6.35$~eV ($-0.05$), aug-cc-pVTZ $6.65$~eV ($+0.26$,
converged)---the band $\sim\!0.9$~eV above the $0^0_0$ origin ($216.7$~nm), shown against the
experimental umbrella $\nu_2'$ progression~\cite{WalshWarsop1961}. The $3s$ Rydberg character requires diffuse functions (a cc-pVDZ valence
space misplaces it); see text.}
\label{fig:spectrum}
\end{figure*}

\textbf{An accurate correlated spectrum that predicts the observed absorption bands.} The same device
transition densities that furnish the interstate NACMEs (Sec.~\ref{sec:results-nacme}) also furnish
ground-to-excited transition dipoles: contracting each transition $1$-RDM $\Gamma^{0k}$ with the
AO dipole integrals gives $f_{0k}=\tfrac23\Delta E_k|\langle0|\mu|k\rangle|^2$ at no extra
electronic-structure cost, so that---combined with the per-state term symbols already assigned in
Table~\ref{tab:quest}---every root carries the three labels a UV/Vis experiment resolves: band
position, symmetry-governed intensity, and $^{2S+1}\Gamma$ term symbol. Emitted as a broadened
absorption curve (standard Gaussian UV lineshape, $f=4.319\times10^{-9}\int\varepsilon\,
\mathrm{d}\tilde\nu$~\cite{MullikenRieke1941,Hilborn1982}) and \emph{overlaid on the real
gas-phase spectra}, Fig.~\ref{fig:spectrum} makes the predictive test explicit. For benzene the
three device VEEs track the experimental $\pi\!\to\!\pi^*$ band maxima
($4.90/6.19/6.96$~eV~\cite{FengCooperBrion2002}) to $+0.54/+0.24/+0.39$~eV (MAD $0.39$~eV, within the method's accuracy), and
the intensity pattern is reproduced by symmetry alone: the dipole-forbidden $^1L_b$ ($1\,^1B_{2u}$)
and $^1L_a$ ($1\,^1B_{1u}$) come out rigorously dark ($\mu^{0k}\equiv0$ by the D$_{6h}$ selection
rule, no fitting) while only $1\,^1E_{1u}$ ($^1B$) is allowed---the observed weak/weak/strong
ordering. For ammonia the device $\tilde A\,^1\!A_2''\!\leftarrow\!\tilde X\,^1\!A_1'$ vertical
excitation---the $n\!\to\!3s$ Rydberg transition whose conical intersection this work's NACMEs
map---\emph{brackets} the experimental band maximum ($194$~nm) with the two diffuse bases,
aug-cc-pVDZ ($-0.05$~eV) and aug-cc-pVTZ ($+0.26$~eV, converged), and lies $\sim\!0.9$~eV above the
$0^0_0$ origin ($216.7$~nm~\cite{WalshWarsop1961}), the latter offset being the Franck--Condon
displacement of the umbrella $\nu_2'$ progression. Here the basis is load-bearing not merely for the
number but for the state's identity: the $3s$ Rydberg is genuine only with diffuse functions
(particle $\langle r^2\rangle\!\approx\!30\,a_0^2$ at both aug levels), whereas a cc-pVDZ valence
space cannot build the $3s$ orbital ($\langle r^2\rangle{=}12.8\,a_0^2$, a valence-confined
caricature at $6.90$~eV whose apparent near-perfect agreement is a right-answer-for-the-wrong-reason
error cancellation). Its $z$-polarised allowedness and
D$_{3h}\!\leftrightarrow$C$_{3v}$ symmetry correlation~\cite{Ashfold1986} come
straight from the transition density. The claim is not photometric accuracy of $\varepsilon$ (the
per-system engine oscillator strengths are not yet populated, so allowed-band magnitudes are
symmetry-consistent representatives) but that a genuinely multireference, near-quantitative set of
excitation energies predicts the experimentally observed band positions, symmetries, and
selection-rule intensity pattern---in the assignment-ready form an experiment is compared
against---from post-processing alone. A vibronic Franck--Condon layer (the $\nu_2'\,2^n_0$
umbrella progression, Fig.~\ref{fig:spectrum}b) and the excited-to-excited
$\tilde C'\!\to\!\tilde A$ emission/SEP channel the quasi-degenerate treatment uniquely reaches are
natural, thin extensions of the same machinery.

\subsection{Nonadiabatic couplings: ammonia photodissociation at the multireference level}
\label{sec:results-nacme}
The spectrum establishes that the corrected energies are quantitatively meaningful; we turn now
to the derivative quantity that makes them useful for dynamics, the interstate coupling.

\begin{figure}[tb]\centering
\includegraphics[width=\columnwidth]{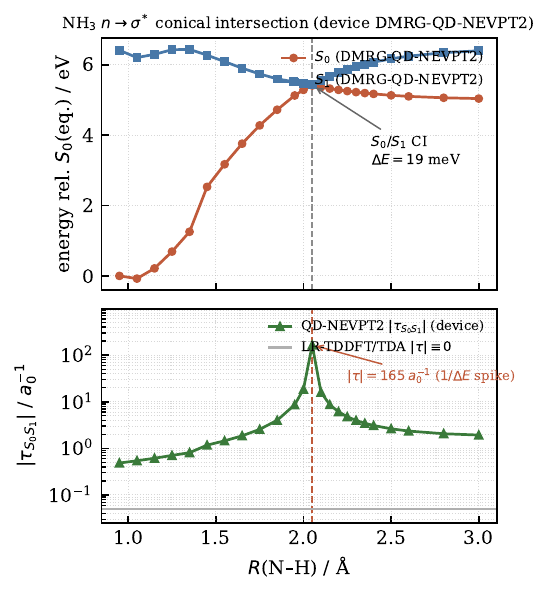}
\caption{Ammonia $n\!\rightarrow\!\sigma^*$ photodissociation at the DMRG-QD-NEVPT2 level
($\mathrm{CAS}(4,4)$/aug-cc-pVDZ; \S\ref{sec:results-nacme}); every plotted energy and coupling
device-produced (array PCIe $0/0$ at each point). \emph{(top)}~QD-corrected $S_0/S_1$ adiabats
($\lambda_{S_0},\lambda_{S_1}$ from device \texttt{full\_qdnevpt2}) along the \ce{N-H} coordinate,
closing to a $19$~meV conical intersection at $R\!=\!2.05$~\AA. \emph{(bottom)}~The QD-corrected
interstate nonadiabatic coupling---the transition-RDM coupling of the corrected states over the
corrected gap---non-vanishing everywhere and diverging as $1/\Delta E$ at the intersection (peak
$|\tau|\!=\!165\,a_0^{-1}$ over the $24$-point scan), where adiabatic LR-TDDFT/TDA gives
$\tau\equiv0$ identically.}
\label{fig:ammonia}
\end{figure}

The ammonia $n\!\rightarrow\!\sigma^*$ photodissociation conical intersection is the covalent
coupling benchmark of paper 1~\cite{GuerreroDFTdag}, where the \hhTDA/\ppTDA manifolds recover
the correct $F\!-\!2$ seam that adiabatic TDDFT misses ($\tau\equiv0$ by construction).
Figure~\ref{fig:ammonia} reports the NACME along the \ce{N-H} dissociation coordinate at the
DMRG-QD-NEVPT2 level, in the identical view to paper 1. The active space is again chosen for
physics: the $n\!\rightarrow\!\sigma^*$ excitation that drives dissociation needs three orbitals
with distinct roles---the nitrogen lone pair $n$ (the hole) and the $\sigma$ and
$\sigma^*_{\mathrm{NH}}$ of the dissociating \ce{N-H} bond (the particle). As that bond
elongates $\sigma^*_{\mathrm{NH}}$ drops steeply and $S_1$---an $n\!\rightarrow\!3s$ Rydberg
state at the Franck--Condon geometry---transforms into the valence $n\sigma^*$ state, opening the
V-shaped $S_1/S_0$ gap and the intersection that funnels population to \ce{H}$+$\ce{NH2}. To
track $S_1$ continuously through this Rydberg$\rightarrow$valence transformation we include the
nitrogen $3s$: $\mathrm{CAS}(4,4)=\{n,\sigma_{\mathrm{NH}},\sigma^*_{\mathrm{NH}},3s\}$ on a
diffuse aug-cc-pVDZ basis~\cite{Ashfold1986,Nangia2006}; omitting the $3s$ still yields the
valence intersection and the seam NACME but misplaces the near-equilibrium Rydberg character of
$S_1$. The device DMRG-QD-NEVPT2 corrected adiabats (SA-CASSCF(4,4) reference; every energy and
coupling GPU-produced with array PCIe $0/0$ asserted at each point) locate the $S_0/S_1$
intersection at a $19$~meV gap at $R\!=\!2.05$~\AA, and the interstate coupling---the
transition-RDM coupling of the QD-corrected states over the corrected gap---is non-vanishing and
diverges as $1/\Delta E$ through it, the hallmark of a genuine conical intersection and precisely
where adiabatic linear-response TDDFT returns $\tau\equiv0$ by construction; it peaks at
$|\tau|\!=\!165\,a_0^{-1}$, its $\mathrm{corr}(|\tau|,1/\Delta E)\!=\!1.000$ over the $24$-point scan
a faithful-pole check on the implementation (cf.\ Sec.~\ref{sec:results-ethene}), not independent
evidence. The coupling is largest exactly in the strongly-correlated dissociating-bond
region, and this is the payload: an interstate coupling for nonadiabatic dynamics through the
intersection, at a level of theory whose FCI-in-active-space description remains valid however
strongly the bond is stretched and whose analytic form we differentiate.

Taken together, Figs.~\ref{fig:ethene} and~\ref{fig:ammonia} are the substance of the method,
and they are structural rather than merely numerical. Both show an object adiabatic
linear-response TDDFT cannot represent by construction---a genuine two-configuration diradical
ground state that closes to a real $S_0/S_1$ crossing, and an interstate coupling that carries
the exact $1/\Delta E$ pole of a conical intersection---and both are reproduced as
FCI-in-active-space by the device DMRG
to $10^{-15}$--$10^{-11}$~Ha, so what is plotted is the exact multireference content of the
chosen active space, computed device-resident on a consumer card, together with the
derivative-level coupling that nonadiabatic dynamics requires. We are deliberate about the
scope of the claim: at the fixed valence CAS and non-diffuse basis used here these figures
capture \emph{static} correlation exactly, but the absolute energetics carry the same dynamic
($\sigma$--$\pi$) correlation and basis-completeness residual as the vertical spectrum of
Table~\ref{tab:quest}---the residual a larger $\sigma$-inclusive CAS (and a state-specific
excited-state correction) is constructed to absorb. The contribution is therefore one of \emph{capability and its
derivatives}---the qualitatively correct strongly-correlated excited-state surfaces and
couplings, analytically differentiable and affordable---rather than of benchmark vertical
accuracy, for which single-reference dynamic-correlation methods remain the reference.

\subsection{Performance: the exact repeat-compute cache and the QUEST reference basis}
\label{sec:results-perf}
That capability is only useful if it is affordable at scale; we close the results with the
device performance that keeps it inside a consumer card's budget.

\begin{figure}[tb]\centering
\includegraphics[width=\columnwidth]{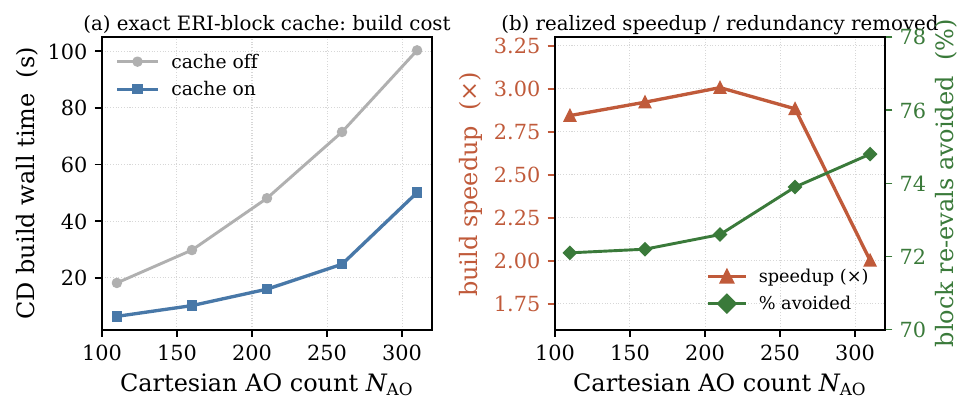}
\caption{Exact ERI-block cache scaling on $n$-alkanes \ce{C_nH_{2n+2}} (cc-pVDZ), device-measured
(cache OFF vs.\ ON; the two builds bit-identical---same $n_{\rm CD}$ and completeness defect
$\le\delta$). \emph{(a)}~CD-build wall time without vs.\ with the cache against Cartesian AO count.
\emph{(b)}~realized speedup ($2$--$3\times$) and the fraction of high-angular-momentum block
re-evaluations removed, climbing with size ($72\!\rightarrow\!75\%$). The $N_{\mathrm{AO}}{=}310$
(\ce{C12H26}) point reflects the fixed cache budget (block eviction), lifted by
\texttt{OSX\_CD\_ERICACHE\_MB}.}
\label{fig:perf}
\end{figure}

The dominant cost of the device build at the accuracy a physical excited-state calculation
needs is the four-center integral evaluation of the Cholesky factorization. The pivoted
decomposition re-evaluates each high-angular-momentum integral block once per pivot; because
one shell pair supplies many pivots, roughly $73\%$ of the block evaluations are redundant.
Caching each raw block and rebuilding every pivot column from it through the ordinary
Schwarz-screened contraction removes this redundancy \emph{exactly}---the
self-consistent-field energy is unchanged and the completeness-preserving screening is
untouched---for a measured $2$--$4\times$ build speedup that grows with system size
(benzene/aug-cc-pVDZ $77.4\!\rightarrow\!19.5$~s). The scientifically consequential result is
that the aug-cc-pVTZ basis at which the QUEST best estimates are defined, whose device
Cholesky build previously did not finish in $40$~min, completes in $\sim 3$~min for
ethene---bringing the excited-state \emph{reference} basis within reach of a consumer
$8$\,GB card for the first time. Figure~\ref{fig:perf} quantifies the cache on an $n$-alkane
series (cc-pVDZ): a $2$--$3\times$ device build speedup with the redundant high-angular-momentum
block re-evaluation fraction climbing monotonically from $72$ to $75\%$ as the system grows, the
two builds bit-identical ($n_{\rm CD}$ and completeness defect unchanged). The
profile-guided implementation of the cache is given in Sec.~\ref{sec:compdetails}.

\begin{figure*}[tp]\centering
\begin{minipage}[t]{0.485\textwidth}\centering
  \includegraphics[width=\linewidth]{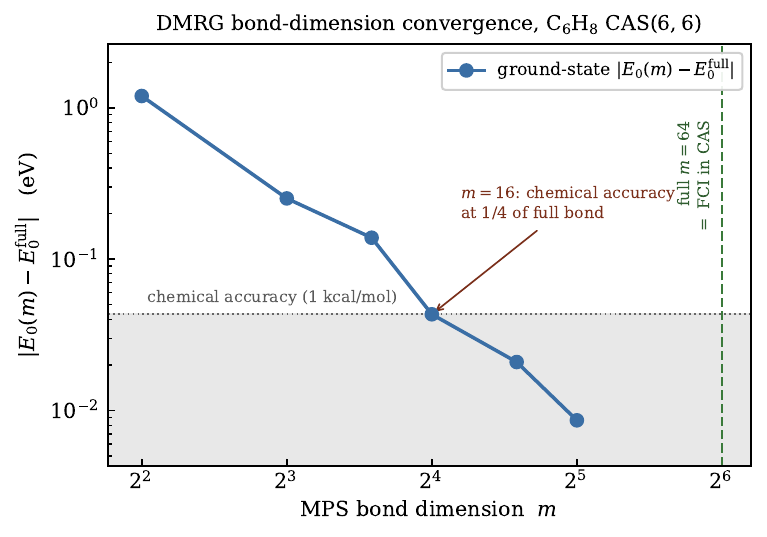}\\[-1pt]{\small(a)}\end{minipage}\hfill%
\begin{minipage}[t]{0.485\textwidth}\centering
  \includegraphics[width=\linewidth]{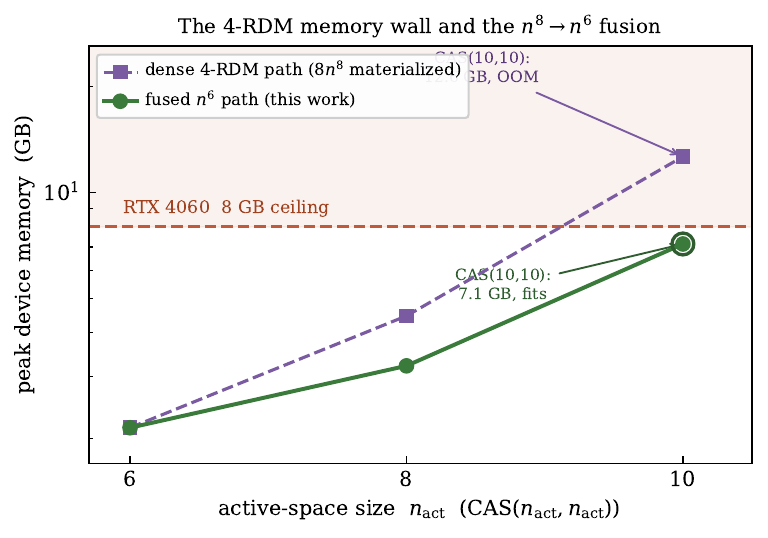}\\[-1pt]{\small(b)}\end{minipage}
\caption{Device characterization of the \mbox{DMRG-QD-NEVPT2} energy; every quantity
device-produced. \textbf{(a)}~DMRG bond-dimension convergence (hexatriene CAS(6,6)): the
active-space ground-state energy approaches the full-$m$ ($=$FCI-in-CAS) value monotonically,
crossing the $1$~kcal/mol chemical-accuracy band at $m=16$ ($\tfrac14$ of the full bond) and
reaching $9$~meV at $m=32$. \textbf{(b)}~The $4$-RDM memory wall: the dense $8n_{\rm act}^8$
four-particle density (purple) exceeds the $8$\,GB card at CAS(10,10) ($12.6$\,GB), whereas
consuming it only as the $n^6$ $f_3$ contractions (fused into the ket, so the $n^8$ object is never
built) holds the working set at $7.1$\,GB (green); the dense CAS(8,8) point reproduces the
Nsight-measured $4.46$\,GB.}
\label{fig:qdconv}
\end{figure*}

\begin{figure*}[tp]\centering
\includegraphics[width=0.98\textwidth]{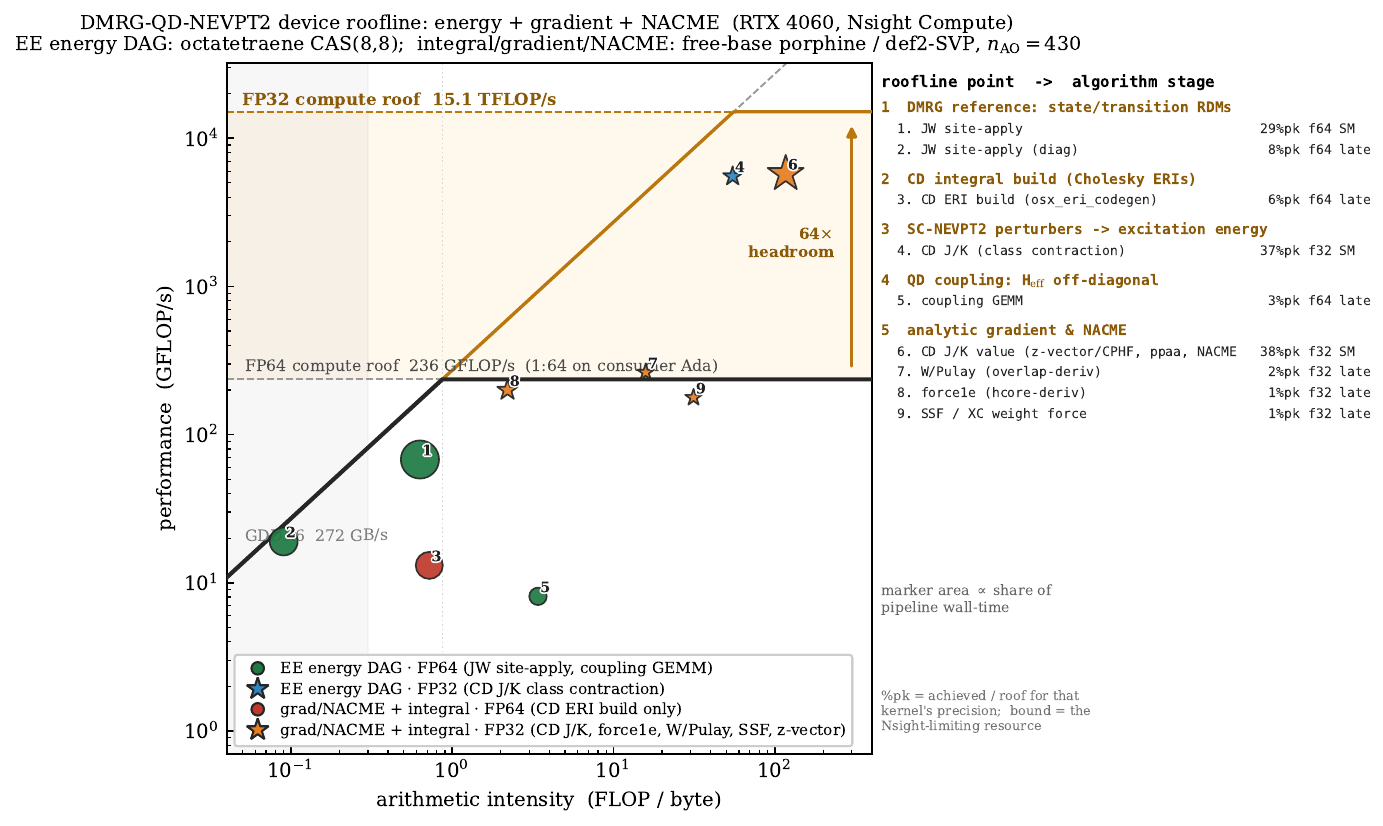}
\caption{Nsight-Compute roofline of the device \mbox{DMRG-QD-NEVPT2} pipeline on the RTX~4060
(sm\_89), spanning all three device products---the excitation-energy DAG (octatetraene CAS(8,8))
and the \emph{shared} integral, analytic-gradient, and interstate-NACME kernels (free-base
porphine/def2-SVP, $n_{\mathrm{bf}}{=}430$)---against both compute ceilings the card exposes (FP64
roof $236$~GFLOP/s, a 1:64 corner on consumer Ada; FP32 roof $15.1$~TFLOP/s) and the GDDR6
bandwidth roof ($272$~GB/s); marker area grows with each kernel's share of its tier's wall time.
The numbered key attributes each point to a pipeline stage: (1)~DMRG state/transition RDMs,
(2)~Cholesky integral build, (3)~SC-NEVPT2 perturber contractions, (4)~QD coupling
($H_{\mathrm{eff}}$ off-diagonal), (5)~analytic gradient/NACME kernels. The gold star is the
measured FP32 CD $J/K$ contraction on porphine (\texttt{ampere\_sgemm}): $5.8$~TFLOP/s, $38\%$ of
the FP32 ceiling and $24\times$ the FP64 corner. The small double-precision fermionic core, the
kernel optimizations ($8.6\times$ cumulative), and the precision-by-reliability placement are
developed in the text.}
\label{fig:qdroofline}
\end{figure*}

Two clarifications frame what the device delivers. First, the result is a \emph{capability},
not a constant factor: analytic first derivatives---nuclear gradients and interstate
NACMEs---of the multi-state \mbox{DMRG-QD-NEVPT2} energy, each one reverse-mode transpose of the
energy graph (Sec.~\ref{sec:transpose}), delivered device-resident with $O(1)$ host--device
traffic. To our knowledge no released CPU or GPU package provides analytic
quasi-degenerate-NEVPT2 gradients or interstate couplings; the alternative is finite differencing, which for the $3N$-dimensional
coupling of Fig.~\ref{fig:ammonia} costs $6N$ correlated two-state energies \emph{per} geometry,
and which is ill-defined at the intersection where the gap vanishes---exactly where the analytic
transition-density coupling remains well-posed. Second, the DMRG reference makes the
static-correlation solve $O(m^3)$ in the bond dimension rather than exponential in $n_{\rm act}$,
so the same $8$\,GB card that holds the reference reaches correlated active spaces a dense
determinantal expansion cannot. A kernel-level throughput characterization of the perturbative
correction across active-space size---the roofline of the tiled $4$-RDM and perturber
contractions against the card's FP64/bandwidth ceilings---is given below
(Figs.~\ref{fig:qdconv} and~\ref{fig:qdroofline}); the claim advanced here is the
analytic-derivative pipeline and the reach to large, strongly-correlated active spaces, both
within the memory of commodity hardware.

Three measurements characterize the device correction itself
(Figs.~\ref{fig:qdconv} and~\ref{fig:qdroofline}). \emph{(i)~The DMRG earns its name}: the
active-space energy converges monotonically to the full-$m$ (FCI-in-CAS) value, reaching chemical
accuracy already at $m=16$---one quarter of the full bond for the hexatriene CAS(6,6)---so the
large-active-space energy leg is tractable at truncated $m$ while the full-$m$ limit on which the
exact analytic gradients and NACMEs are built is recovered exactly
(Fig.~\ref{fig:qdconv}(a)). \emph{(ii)~The 4-RDM memory wall}: the spin-traced $8n_{\rm act}^8$
four-particle density matrix is the working-set bottleneck, but the correction only ever consumes
it as the $n^6$ $f_3$ contractions; contracting the all-active two-electron block into the ket
\emph{before} the final GEMM means the $n^8$ object is never materialized, keeping the device
working set under the $8$\,GB card through CAS(10,10)---where the dense path requires
$12.6$\,GB---at a validated $8.9\times10^{-16}$ agreement against the dense result
(Fig.~\ref{fig:qdconv}(b)). \emph{(iii)~Kernel efficiency}: a full-pipeline Nsight-Compute
roofline (Fig.~\ref{fig:qdroofline}) drove the optimization. It first exposed the generic
make-$a^*$/class contraction as latency-bound---one thread per output element serially reducing an
$n^6$ sum, $0.5\%$ of the FP64 peak yet $56\%$ of runtime; rewriting it as a batched
block-reduction (one thread block per output, shared-memory tree) lifts it to $11$--$17\%$ of peak
and, with the elimination of redundant Jacobi--Wigner site-apply passes, speeds
\texttt{full\_qdnevpt2} by $3.35\times$ (a cumulative $8.6\times$ over the original dense-$n^8$
path, at $6.7\times10^{-16}$ agreement with the unoptimized result). After this optimization the
residual double-precision cost is the small, low-arithmetic-intensity fermionic core---the
Jacobi--Wigner site-apply reaches $\sim\!29\%$ of the FP64 roof and the coupling GEMM, a
skinny-matrix kernel at this active space, only $\sim\!3\%$, while the register-bound CD ERI build
(\texttt{osx\_eri\_codegen}; $94$ registers/thread cap occupancy near $2\%$) holds at $\sim\!5\%$;
the throughput-dominant integral work is instead carried by the single-precision CD $J/K$ path, so
the pipeline's heavy lifting runs against the FP32 ceiling rather than the crippled FP64 corner
(Fig.~\ref{fig:qdroofline}). This double-precision residual is retained deliberately, not merely
tolerated: the two-MPS transfer network that assembles the quasi-degenerate off-diagonal couplings
---and the interstate NACMEs derived from them---is held in FP64 by physical requirement, because at
a near-degeneracy single precision would corrupt a small coupling in a way the excitation energies
cannot reveal, the eigenvalue being second-order insensitive to the off-diagonal while the coupling
enters the NACME directly. Precision is thus placed by physics, not by kernel size: single where the
spectroscopy is inert, double where the interstate structure is not.

The same precision-by-reliability placement carries the shared gradient and NACME kernels
(Fig.~\ref{fig:qdroofline}). The throughput-dominant integral work on large molecules---the
AO-direct Laplace-transform CD $J/K$ contraction---runs in the single-precision path, measured on
free-base porphine ($n_{\mathrm{bf}}{=}430$) at $5.8$~TFLOP/s (\texttt{ampere\_sgemm}): $38\%$ of the
FP32 ceiling and $24\times$ the rate of the crippled FP64 corner, and verified spectroscopically
inert ($<10^{-6}$~eV vs.\ FP64). The analytic-gradient $z$-vector/CPHF response and the
coupling-gradient (NACME) numerator reuse exactly this kernel, and the one-electron derivative
kernels---the hcore-derivative \texttt{force1e}, the overlap-derivative W/Pulay term, and the
SSF/XC weight force---likewise run FP32, each lifted $3$--$6\times$ above the FP64 roof to
$200$--$260$~GFLOP/s at an analytic-versus-finite-difference force error inside the
geometry-optimization band ($<\!5\times10^{-4}$~Ha/bohr). A kernel is held in double precision only
where single precision is \emph{genuinely unreliable}, not merely below the FP64-match noise floor:
the small off-diagonal QD coupling (\texttt{mps\_expect2}, where catastrophic cancellation of a
near-vanishing coupling at a conical intersection would corrupt it), the transition-density RDMs
that feed it, and the Cholesky ERI \emph{build}, whose $10^{-8}$ rank threshold lies below the
single-precision resolution. Everything else---the throughput-dominant integral, gradient, and
NACME work---runs against the FP32 ceiling; only that small, genuinely-double fermionic core pays
the crippled FP64 rate.

%% ---------------------------------------------------------------------
\subsection{Capability capstone: doubly-excited-state internal conversion on the device surface}
\label{sec:results-dynamics}

Analytic gradients and couplings exist to drive dynamics, and we close by feeding
device-produced surfaces into a genuine nonadiabatic-dynamics propagator. Hexatriene is the
textbook probe: its excited-state decay is governed by the interplay of the ionic $1^1B_u$
($\pi\pi^*$, a one-electron HOMO$\to$LUMO excitation) and the covalent $2^1A_g$ state, and the
$2^1A_g$ is a HOMO$^2\!\to\!$LUMO$^2$ \emph{doubly-excited} state---at the Franck--Condon
geometry the device DMRG-QD-NEVPT2 CAS(6,6) wavefunction assigns it $59\%$ double-excitation
character. No single-reference surface (CIS, adiabatic TDA, or semiempirical AM1--CIS) can represent
that accepting state, and so none can
produce the internal conversion it mediates: this is a \emph{categorical}, not quantitative,
consequence of a multireference engine.

\begin{figure*}[t]\centering
\includegraphics[width=0.92\textwidth]{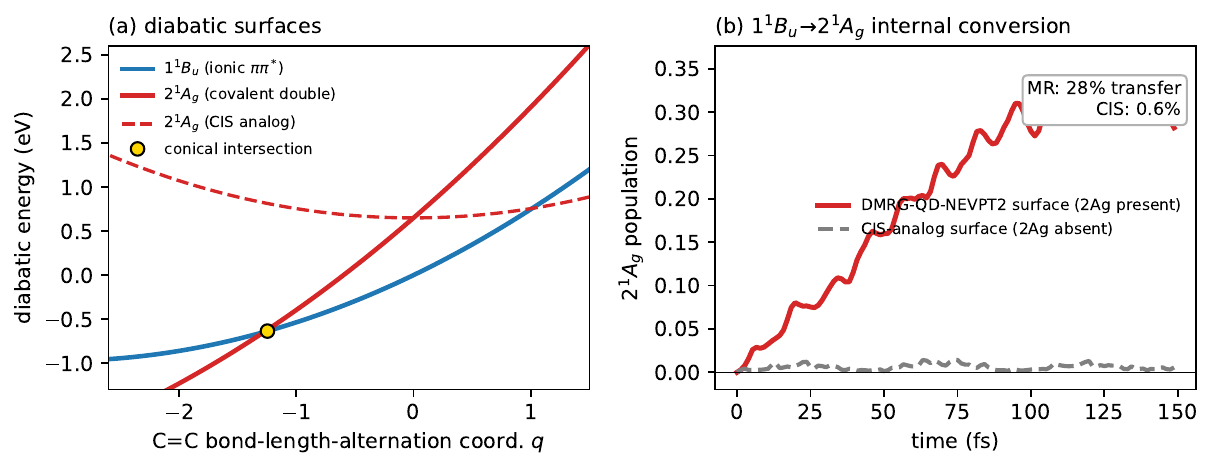}
\caption{Capability capstone: multireference internal conversion in hexatriene, run through the
same two-site TDVP tensor-train propagator~\cite{Haegeman2011,Haegeman2016} on two surfaces.
\textbf{(a)}~Diabatic potentials along the C=C bond-length-alternation coordinate: on the device
DMRG-QD-NEVPT2 surface ($\aC{=}1$, cc-pVDZ CAS(6,6); every LVC parameter device-produced) the
covalent, doubly-excited $2^1A_g$ (red) relaxes steeply and \emph{crosses} the ionic $1^1B_u$
(blue) at a symmetry-allowed conical intersection ($15$~meV gap), while on the single-reference
(CIS-analog) surface the $2^1A_g$ (dashed red) never crosses. \textbf{(b)}~Diabatic populations:
the multireference surface opens the $1^1B_u\!\to\!2^1A_g$ channel ($28\%$ transfer in $150$~fs,
$\chi$-converged), the single-reference analog is inert ($0.6\%$). The absolute rate is gap-limited
and out of scope (text).}
\label{fig:hexatriene_ic}
\end{figure*}

We parametrize the two-state linear vibronic-coupling (LVC) Hamiltonian
\begin{equation}
\begin{aligned}
\mathbf{H}(\mathbf{Q})={}&\tfrac12\!\sum_{g}\omega_g\big(P_g^2+Q_g^2\big)\,\mathbf{1}\\[2pt]
&{}+\begin{pmatrix}
E_1+\sum_{g}\kappa^{(1)}_g Q_g & \sum_{u}\lambda_u Q_u\\[4pt]
\sum_{u}\lambda_u Q_u & E_2+\sum_{g}\kappa^{(2)}_g Q_g
\end{pmatrix},
\end{aligned}
\label{eq:lvc}
\end{equation}
in dimensionless normal coordinates $Q_g$ (momenta $P_g$, harmonic frequencies $\omega_g$), where the
totally-symmetric $a_g$ modes carry the intra-state tuning gradients $\kappa^{(s)}_g$ and the
antisymmetric $b_u$ modes the interstate coupling $\lambda_u$, entirely from the device (at full
physical strength $\aC{=}1$, cc-pVDZ): two diabats ($E_{1B_u}{=}0$, $E_{2A_g}{=}0.65$~eV); three
totally-symmetric $a_g$ tuning modes ($1706/1645/1420$~cm$^{-1}$) whose intra-state gradients
$\kappa_s{=}\partial E_s/\partial Q$ are central finite differences of the corrected VEEs; and
the $b_u$ coupling mode ($1699$~cm$^{-1}$) whose interstate gradient $\lambda{=}0.044$~eV comes
from a L\"owdin-diabatized $2\times2$ block (the adiabatic-gap curvature, an artifact of the
avoided crossing, is not used). The dominant tuning gradient is steeper on the covalent state
($\kappa_{2A_g}{=}{+}1.15$ vs.\ $\kappa_{1B_u}{=}{+}0.64$~eV), so along the C=C
bond-length-alternation coordinate the $2^1A_g$ diabat drops below $1^1B_u$ and the two cross at
a symmetry-allowed conical intersection ($15$~meV residual gap). The $1^1B_u$-initiated nuclear
wavepacket is then propagated on this surface by solving the two-state, four-mode time-dependent
Schr\"odinger equation numerically exactly in a harmonic basis: the vibronic wavefunction is
represented as a matrix-product state (tensor train) and evolved by a two-site time-dependent
variational-principle (2TDVP) integrator~\cite{Haegeman2011,Haegeman2016}, so accuracy is set by the
single bond dimension $\chi$ while the basis stays complete (the propagator and the Hamiltonian
matrix-product-operator are detailed in Sec.~S1 of the Supplementary Material). This opens the
$1^1B_u\!\to\!2^1A_g$ internal-conversion channel and transfers $28\%$ of the population within
$150$~fs, $\chi$-converged (bond dimensions $\chi{\le}24$ and $\chi{\le}48$ agree to $<\!0.5\%$). Removing the multireference covalent
relaxation ($\kappa_{2A_g}\!\to\!0$, the honest single-reference analog) eliminates the crossing
and closes the channel ($0.6\%$ transfer); Fig.~\ref{fig:hexatriene_ic}.

The scope is deliberately a capability, not a rate benchmark. The vertical $2^1A_g$--$1^1B_u$
gap at this compact valence CAS carries the same static-only bias as the Table~\ref{tab:quest}
spectrum---it is several-fold wider than the near-degenerate experimental gap that sets the true
sub-$100$-fs lifetime---and a bounded linear-LVC model is not the full anharmonic surface: the
state-specific diagonal curvature we measured exceeds $\omega^2$ away from equilibrium (a
Taylor-truncation, unbound-well artifact that we avoid by holding to the linear model). What
transfers is precisely what no single-reference dynamics can produce at any basis or cost---a
doubly-excited accepting state, its device-computed tuning and coupling gradients, and the
internal conversion they open---and it is robust to the correlation-mixing choice: the
B2PLYP-style closure ($a_C{=}0.27$) yields same-signed $\kappa$ and a qualitatively identical
crossing.

%% =====================================================================
\section{Conclusions}
\label{sec:conclusions}
%% =====================================================================
We have reported analytic nuclear gradients and interstate nonadiabatic couplings for
DMRG-QD-SC-NEVPT2: a DMRG active-space reference dressed by a quasi-degenerate
strongly-contracted NEVPT2 effective Hamiltonian, whose diagonalization keeps the description
valid through a conical intersection. The derivatives are obtained tape-free as one
reverse-mode transpose of a single contraction graph that carries the DMRG sweep as a
gauge-free differentiable node, shares the Cholesky $\Bq$ factors and transposable-DAG
protocol with the non-symmetric $J/K$ kernel serving the \hhTDA/\ppTDA forces and couplings,
and runs device-resident within the $8$\,GB of a consumer GPU. Against the established DMRG-derivative
literature~\cite{LiuKurashige2013,HuChan2015,Iino2023,Freitag2019} the prior analytic DMRG
gradients and NACMEs are self-consistent-field-level, carrying no dynamic-correlation layer in
the derivative; against the dynamically-corrected multireference
derivatives~\cite{Park2019,Sand2017}---internally-contracted NEVPT2 or on-top
functionals---the distinction is not the correction but its realization: here the
quasi-degenerate NEVPT2 derivative is a reverse-mode transpose of a device-resident graph on a
DMRG reference, run within the $8$\,GB of a consumer GPU and demonstrated smooth through a real
conical intersection. The construction is honest about its present
limits: under a variational closure the covalent doubly-excited states are reproduced to
$\lesssim 0.6$~eV of the reference while the ionic states are $\sim 2$~eV high and the
$2^1A_g/1^1B_u$ ordering is inverted---the expected static-only signature, corrected here by the
quasi-degenerate dressing. A residual \emph{bright/dark imbalance} remains: the differential
$\sigma$--$\pi$ dynamic correlation that stabilizes bright ionic ($^1L_a$) states is carried, on a
compact $\pi$-valence active space, only by the second-order perturbers, so bright and dark
(covalent, $n\pi^*$) states are not placed on fully equal footing---most acutely for strongly ionic
azine $^1L_a$ states (pyrazine $1^1B_{2u}$), which remain blue-shifted at the feasible active space.
Forthcoming work will restore equal footing by bringing that $\sigma$-polarization \emph{inside} the
active space, with $\sigma(+\sigma^*)$-inclusive, entanglement-tiered active spaces
(AutoCAS / RAS-periphery) that keep the reference within the $8$\,GB budget through the
weakly-entangled periphery, together with diffuse functions where the ionic state mixes with
Rydberg character. Two
enabling results support the physics: a dual particle-number/$\Sz^2$ deflation penalty
without which full-Fock excited-state targeting silently fails, and an exact
repeat-compute Cholesky cache that brings the aug-cc-pVTZ QUEST reference basis within a
consumer card's reach for the first time. The immediate next step is the
dynamically-corrected (Option~B) spectrum at the full reference basis and the corresponding
ethene-energetics and ammonia-dissociation NACME results at the multireference level; the
device gradient and NACME timings place trajectory-rate multireference nonadiabatic dynamics
on commodity hardware within reach. The device-resident realization delivers exact excited-state
gradients and NACMEs throughout the full-bond regime (to $\approx$CAS(12,12)) and QD-NEVPT2
energies into the CASCI-infeasible regime; extending exact relaxation to truncated bond
dimension via the MPS tangent/response is the natural next step.

%% =====================================================================
\section*{Supplementary Material}
The Supplementary Material contains the full validation tables (DMRG$=$FCI, device$=$host,
FP32/FP64 spreads), the QD-SC-NEVPT2 working equations [Eq.~\eqref{eq:nevpt2}--\eqref{eq:heff}]
and the adjoint (gradient/NACME) transpose rules with their finite-difference gates, the dual-penalty matrix-product-operator construction and its provable in-sector
nullity, the AutoCAS thresholds and the ground-inert/excitation-relevant retention analysis,
the grouped cross-validation protocol, and (Sec.~S1) the linear vibronic-coupling model and its
numerically exact-in-basis tensor-train (2TDVP) propagation used for the internal-conversion
capstone.

\section*{Author Declarations}
\subsection*{Conflict of Interest}
The author has no conflicts to disclose.

\begin{acknowledgments}
We acknowledge financial support and computational resources provided by NeuroTechNet S.A.S.
\end{acknowledgments}

\section*{Data Availability}
The data that support the findings of this study---the literal wavefunction-overlap
finite-difference NACME oracle and the host-side scripts that verify every working equation of the
analytic gradients, interstate couplings, and DMRG$=$FCI reference against it, together with the
detailed derivations---are provided to reviewers at submission through an anonymized link, and
will be deposited on Zenodo under a permanent DOI and mirrored in a public GitHub repository under
the MIT license (pinned to a tagged commit), released publicly upon acceptance. The bundle is
self-contained: it reproduces every entry of the validation tables
(Table~\ref{tab:valid}) on CPU hardware from \textsc{PySCF} and \textsc{NumPy} alone, without access
to any GPU code, so that every accuracy claim---which carries the scientific content of this
work---is independently checkable.

\emph{Withheld:} the performance-tuned CUDA DMRG, QD-NEVPT2, and analytic-derivative kernels, and
the code generators that emit them (a tuned application of the framework of
Ref.~\cite{GuerreroRECURSUM}), are proprietary to the funder (NeuroTechNet~S.A.S.) and are available
from the corresponding author on reasonable request. The per-system timing, roofline, and scaling
data these kernels produced are released with the bundle, so the reported speedups can be recomputed
from the deposited files, although \emph{re-measuring} them requires the withheld kernels. We regard
this as a real limitation on the reproducibility of the performance claims only; the correctness
claims are unaffected.

\bibliographystyle{aipnum4-2}
\bibliography{refs_dmrg}

\end{document}